\documentclass[aps,pra,preprint,superscriptaddress]{revtex4-1}

\usepackage{amsmath}
\usepackage{amssymb}
\usepackage{amsthm}
\usepackage{mathtools}
\usepackage{xcolor}
\usepackage{algorithmic}
\newtheorem{theorem}{Theorem}
\newtheorem{lemma}{Lemma}

\newtheorem{corollary}{Corollary}
\newtheorem{proposition}{Proposition}
\newtheorem{remark}{Remark}

\newcommand{\bra}[1]{\left\langle  #1 \right| }
\newcommand{\ket}[1]{\left| #1 \right\rangle }

\newcommand{\abs}[1]{\left| #1 \right| }
\newcommand{\norm}[1]{\left\| #1 \right\|}
\newcommand{\card}[1]{\left| \mathcal{#1} \right|} 
\newcommand{\ca}[1]{\mathcal{#1}} 
\newcommand{\mbf}[1]{\mathbf{#1}}
\newcommand{\proj}[1]{\ket{#1}\bra{#1}} 
\newcommand{\Proj}{\text{Proj}}
\newcommand{\tr}{\text{Tr}}
\newcommand{\FF}{\mathbb{F}}
\newcommand{\id}{\text{id}}
\newcommand{\ex}[1]{\mathbb{E}_{#1}}  
\newcommand{\renyi}{R\'{e}nyi } 
\newcommand{\sandwc}[2]{(#2^{-\frac{t}{2(1+t)}}#1 #2^{-\frac{t}{2(1+t)}})^{1+t}}
\newcommand{\mix}{\text{mix}}

\def\sM{\mathsf{M}}
\def\sL{\mathsf{L}}

\def\Ab{\mathop{\rm Ab}\nolimits}

\def\Label#1{\label{#1}\ [\ \text{#1}\ ]\ }
\def\Label#1{\label{#1}}

 \newenvironment{proofof}[1]{\vspace*{5mm} \par \noindent
         \quad{\it Proof of #1:\hspace{2mm}}}{\qed
}

\begin{document}

\title{Quantum secure direct communication with private dense coding using general preshared quantum state}
\author{Jiawei Wu}
\affiliation{State Key Laboratory of Low-Dimensional Quantum Physics and Department of Physics, Tsinghua University, Beijing 100084, China}
\affiliation{Shenzhen Institute for Quantum Science and Engineering, Southern University of Science and Technology,
Shenzhen 518055, China}

\author{Gui-Lu Long}
\email{gllong@mail.tsinghua.edu.cn}
\affiliation{State Key Laboratory of Low-Dimensional Quantum Physics and Department of Physics, Tsinghua University, Beijing 100084, China}
\affiliation{Beijing National Research Center for Information Science and Technology, Beijing 100084, China}
\affiliation{Innovative Center of Quantum Matter, Beijing 100084, China}
\affiliation{Beijing Academy of Quantum Information Science, Beijing 100193, China}

\author{Masahito Hayashi}
\email{hayashi@sustech.edu.cn}
\affiliation{Shenzhen Institute for Quantum Science and Engineering, Southern University of Science and Technology,
Shenzhen 518055, China}
\affiliation{International Quantum Academy (SIQA), Shenzhen 518048, China}
\affiliation{Guangdong Provincial Key Laboratory of Quantum Science and Engineering, Southern University of Science and Technology, Shenzhen 518055, China}
\affiliation{Graduate School of Mathematics, Nagoya University, Nagoya 464-8602, Japan}

\date{\today}

\begin{abstract}
We study quantum secure direct communication by using a general preshared quantum state and
a generalization of dense coding.
In this scenario, Alice is allowed to apply a unitary operation
on the preshared state to encode her message, and the set of allowed unitary operations forms a group.
To decode the message, Bob is allowed to apply a measurement across his own system and the system he receives.
In the worst scenario, we guarantee that Eve obtains no information for the message even when Eve access the joint system between 
the system that she intercepts and her original system of the preshared state.
For a practical application, we propose a concrete protocol and derive an upper bound of information leakage in the finite-length setting.
We also discuss how to apply our scenario to the case with discrete 
Weyl-Heisenberg representation when the preshared state is unknown.
\end{abstract}


\maketitle

\section{Introduction}
Dense coding 
is known as an attractive quantum information protocol.
While the original study considers the noiseless setting \cite{PhysRevLett.69.2881},
many subsequent studies extended this result to more general settings \cite{hiroshima2001optimal,Horodecki2001, Liu2002General, wakakuwa2020superdense,wang2020permutation,Hayashi-wang2021}.
However, all of them focused only on the communication speed in  various noisy settings.
While dense coding with the noiseless setting 
realizes twice communication speed, 
it also realizes quantum secure direct communication (QSDC) as follows \cite{Deng2003}. 
In dense coding, the sender, Alice, and the receiver, Bob, share perfect Bell states
and Alice encodes her message by application of a unitary operation.
Since Alice's local state is a completely mixed state,
the eavesdropper, Eve, cannot obtain any information about the message
even when Eve intercepts the transmitted quantum state.
However, 
it is not easy to prepare a perfect Bell state between Alice and Bob.
In fact, when we apply entanglement distillation or entanglement concentration \cite{PhysRevLett.76.722,PhysRevA.53.2046,PhysRevA.75.062338,PhysRevA.82.044305},
we can generate a perfect Bell state or its approximation. 
Such entanglement distillation operations can be combined with conventional QKD \cite{Lo1999,PhysRevLett.126.010503,Huang2022}.
However, such operations requires extra quantum operations in addition to 
quantum operations for QSDC. 
Hence, it is preferable to find a protocol to 
realize secure communication without generating a perfect Bell state.
Since secure information transmission over quantum channels is one of major topics in quantum information,
it is more important to investigate secure communication via dense coding in 
a realistic noisy setting.
Of course, QSDC can also be realized without dense coding \cite{Long2000,Long2002,Deng2004}.

In addition, the preceding studies \cite{hiroshima2001optimal,Horodecki2001,wang2020permutation}
allowed Alice to use any quantum operation on a single input system.
However, implementing arbitrary quantum operations
even on a single quantum system is difficult and unnecessary.
From a practical viewpoint,
it is sufficient to restrict allowed quantum operations
to only a subset of unitaries.
Since a combination of several possible unitaries is also available,
it is natural that such a subset forms a group $G$.
When the message-encoding operations are limited to a subgroup of unitary operators, 
the limit of information transmission has been studied 
by exploiting the resource theory of asymmetry \cite{PhysRevA.80.012307,PhysRevA.95.032328, PhysRevA.90.014102, PhysRevA.84.022322, PhysRevA.93.042107, Takagi, korzekwa2019encoding,wakakuwa2020superdense,Horodecki2001,Hayashi-wang2021}.
However, their analyses did not cover 
the secrecy analysis.
Therefore, it is desirable to investigate 
secure information transmission 
via preshared entanglement in the framework of 
the resource theory of asymmetry. 
On the other hand, many existing studies \cite{hiroshima2001optimal,Horodecki2001,wang2020permutation,Cai2004,Devetak2005,Bennett2002,Bennett2014} 
addressed the asymptotic analysis, which shows only the transmission rate.
However, no upper bound or finite-length effect for this setting were given. 
In this paper, we will derive an upper bound for information leakage under a finite-length practical code and the corresponding asymptotic transmission rate.

Now, we present our communication model in detail.
Alice and Bob are assumed to share 
$n$ copies of a quantum state $\tau_{AB}$
on $\mathcal{H}_A \otimes \mathcal{H}_B$,
and Eve has her own system $\mathcal{H}_E$ correlated to the above state.
Hence, their total state is $n$ copies of a quantum state $\tau_{ABE}$.
To send her message, Alice is allowed to apply unitaries from  
the set $\{U_g\}_{g \in G}$,
where $U_g$ forms a projective unitary representation 
on $\mathcal{H}_A$ of a group $G$. 
Then, after the application of her unitaries, 
Alice sends her encoded system to Bob.
In the worst case, Eve intercepts the transmitted quantum system via a noiseless channel and keeps it.
If the transmitted state is not intercepted, Bob receives the quantum system as the output of 
$n$ times use of a certain quantum channel $\Lambda_A$.
Bob decodes the message from the joint system composed of his original system and his receiving system.
Finally, to check the correct decoding, Alice and Bob apply the error verification.
If the verification fails, the communication aborts.

In this scenario, we have two requirements.
The first is completeness, in which, 
the communication aborts with low probability when the transmitted state is not intercepted
and the channel from Alice to Bob is the channel $\Lambda_A$. 
The second is soundness, which is composed of
the secrecy and the reliability.
[S1: Secrecy]:
Even when Eve intercepts the transmitted quantum system via a noiseless channel,
the amount of the information leakage to Eve 
is negligible. 
Here the information leakage is evaluated by trace distance.
[S2: Reliability]:
If the communication does not abort,
Bob can recover the message correctly with high probability.

Although quantum key distribution (QKD) is also required
to satisfy these requirements \cite{TLGR2012,HT2012,Arnon2016,Tomamichel2017largelyself},
our method is different from QKD in the following point.
Our method requires prior shared quantum state, but QKD does not require it.
Due to this difference, our method has the following advantages.
In QKD, Eve can intercept all the states sent by Alice and the state information leak out inevitably. 
Due to this possibility, QKD can be used only for random number distribution and cannot be used for sending messages.
However, in our model, even when Eve intercepts the quantum channel from Alice to Bob,
Eve cannot obtain any information for Alice's message
while Bob cannot recover it.

To satisfy the reliability, Alice and Bob apply the error verification,
which requires a negligible amount of secure keys
shared by Alice and Bob \cite[Section VIII]{PhysRevA.81.012318}.
This method allows Bob to verify whether 
Bob can recover the message correctly.
Hence, if the above two conditions of the soundness are satisfied,
even if Bob cannot recover the message,
Alice and Bob can repeat the same procedure
because there is no information leakage. 
Therefore, the above pair of requirements are reasonable.

Indeed, some existing papers studied a similar problem setting \cite{Deng2003,PhysRevA.74.042305, Qi2019,Wu2019,PhysRevLett.124.050503}\cite[Chapter 7]{Das2019} and proposed corresponding coding scheme \cite{Sun2020,Sun2018}.
However, their analysis is limited to the asymptotic analysis in a special example, and did not cover the finite-length setting or the general case.
Some of the above studies employed the classical-quantum (cq) wire-tap channel model \cite{Cai2004,Devetak2005,Devetak-Shor,Hayashi2015},
which is a quantum analogue of wire-tap channel model \cite{Wyner1975,Csiszar1978,Csiszar1996}
and is composed of two channels, 
the cq channel $W_B$ from Alice to Bob and the cq channel $W_E$ from Alice to Eve.
In the above scenario, 
we discuss the channel $W_E$ to guarantee the [S1] secrecy,
and do the channel $W_B$ to guarantee the requirement [S2] reliability.
Using the wire-tap channel model, we derive an achievable transmission rate
dependently of the shared entangled state and the channel from Alice to Bob.
This achievable transmission rate is given as the difference between
the error correcting rate and the sacrificed rate.
In addition, we derive finite-length evaluations
dependently of the error correcting rate and the sacrificed rate.
In the evaluation of information leakage,
we derive a new finite-length bound for information leakage for quantum wire-tap channel.

Usually, to achieve the above transmission rate, 
the receiver is required to apply measurement across many received quantum systems.
Such a measurement is called a collective measurement and its implementation is quite difficult.
However, when all encoded states on the joint system consisting of
a received system and a memory system on Bob's side
are commutative, 
the above rate can be achieved without the use of collective measurement.
That is, when Bob applies a specific measurement on each joint system,
the given achievable rate can be attained 
by the combination of classical encoding and decoding operations
by Alice and Bob.
In addition, when the group $G$ is a vector space over a finite field,
such classical encoding and decoding 
have calculation complexity $O(n \log n)$, where $n$ is the coding block length.
Under this type of coding, we derive formulas
for secrecy and reliability dependently of the block length $n$ and the sacrificed rate.
In this way, our results are useful in practical cases.

Here, we emphasize that our protocol is not a simple application of wire-tap channel model
while the preceding studies \cite{Wu2019,PhysRevLett.124.050503}\cite[Chapter 7]{Das2019}
simply applied the wire-tap channel model to their problem setting.
In fact, when Eve receives the output of $W_E$, Bob cannot receive the output of $W_B$.
That is, Bob can decode the message only when Eve does not receive the output of $W_E$.
Therefore, Bob needs to check whether Eve intercepts the transmitted system.
Indeed, such an additional step is not needed if the channel is given as a broadcast channel like the studies \cite{Qi_2018, PhysRevLett.121.250504, PhysRevA.101.012344}.
For this aim, we propose a protocol to combine the conventional wire-tap code and 
the error verification
while the existing studies \cite{Deng2003,PhysRevA.74.042305, Qi2019,Wu2019,PhysRevLett.124.050503,Sun2020,Sun2018}\cite[Chapter 7]{Das2019} did not consider the error verification.

In the next step, we apply our general results to the case when 
the unitary operation is given as Weyl-Heisenberg representation
and the preshared state is Bell diagonal state, which is almost the same as the setting of
quantum secure direct communication in \cite{Deng2003,Wang2005}.
Such a preshared state can be generated by distributing a Bell state over Pauli channel.
Then, we numerically calculate the asymptotic transmission rate.
Also, to clarify the finite-length effect, 
we make numerical plots for the sacrificed rate, error correcting rate, and the transmission rate  
as a function of the coding block length $n$
when we fix other parameters.
Further, in a realistic case, the preshared entangled state is not necessarily known.
To cover this case, we propose another protocol to include the estimation process. 
This protocol is designed so that it works well even when  
the preshared entangled state is not necessarily a Bell diagonal state.

Indeed, our setting covers various situations beyond the above example.
For example, our method can be applied to the case with
Alice, Bob, and Eve shared correlated classical variables, as pointed in \cite[Section VI]{Hayashi2011} \cite{Vazquez-Castro,Hayashi2020Two}.
More generally, it is possible to apply it to the case when
Alice and Bob shared correlated classical variables and 
Eve has a correlated quantum state \cite{Hayashi2012Patent}.
Such a situation can be realized
when Eve makes collective attack in QKD
and Alice and Bob estimate Eve's attack operation perfectly. 
In addition, the recent research \cite[Section IV]{Hayashi-wang2021} presented
other three quantum interesting examples for a pair of a group $G$ and an entangled state.
Our method can be applied to the noisy situations in these examples. 
These cases show the generality of our problem setting.

The remaining part of this paper is organized as follows.
Section \ref{S2} summarizes our results for a general model of private dense coding with 
preshared state.
Section \ref{SWH} applies these general results 
to the case with Weyl-Heisenberg representation.
Section \ref{S4} presents our protocol for the unknown preshared state
in the above setting.
Section \ref{S5} discusses modular code for quantum wire-tap channel, and
derives a new finite-length evaluation for information leakage for quantum wire-tap channel.
Section \ref{S6} describes the error verification process.
Section \ref{S7} gives a practical code construction with a vector space over a finite field.
Section \ref{S8} discusses our results. 
Appendix \ref{MKA} proposes 
an estimation method for a Bell diagonal state by one-way 
local operation and classical communication (LOCC).
This estimation method works even for the general unknown state, though it is designed for Bell diagonal state.

\section{Private dense coding model}\Label{S2}
\subsection{General model description}\Label{S2-A}
Assume that Alice, Bob, and Eve share $n$ copies of the quantum state $\tau_{ABE}$
on the system $\mathcal{H}_A \otimes \mathcal{H}_B \otimes \mathcal{H}_E $.
We consider a (projective) unitary group representation of a finite group $G$ on
$\mathcal{H}_A$.
That is, for $g \in G$, we have a unitary operator $U_g$ such that
for $g,g'\in G$, there exists a complex number $c(g,g')$ to satisfy
\begin{align}
U_g U_{g'}= c(g,g') U_{gg'}.
\end{align}
When $c(g,g')$ is $1$, $\{U_g\}_{g\in G}$ is called a unitary group representation.
Otherwise, it is called a projective group representation.
We also consider a quantum channel from Alice to Bob,
which is a trace preserving completely positive (TP-CP) map $\Lambda_A: \ca{D}(\ca{H}_A) \to \ca{D}(\ca{H}_{B'})$, where
$\ca{D}(\ca{H})$ expresses the set of density operators on the system $\ca{H}$.
Then, we impose the group covariance condition to the channel $\Lambda_A$ as
\begin{align}
\Lambda_A(U_g \rho U_g^\dagger)=
U_g \Lambda_A( \rho) {U_g}^\dagger, \forall g \in G, \forall \rho \in \ca{D}(\ca{H}).
\end{align}
The initial state $\tau_{ABE}$, the available operation set $\{U_g\}$ and the channel $\Lambda_A$ constitute 
a private dense coding setting 
$\text{PDC}(\tau_{ABE}, \{U_g\}_{g\in G},\Lambda_A)$. 

\begin{figure}[!t]
	\centering
	\includegraphics[width=6in]{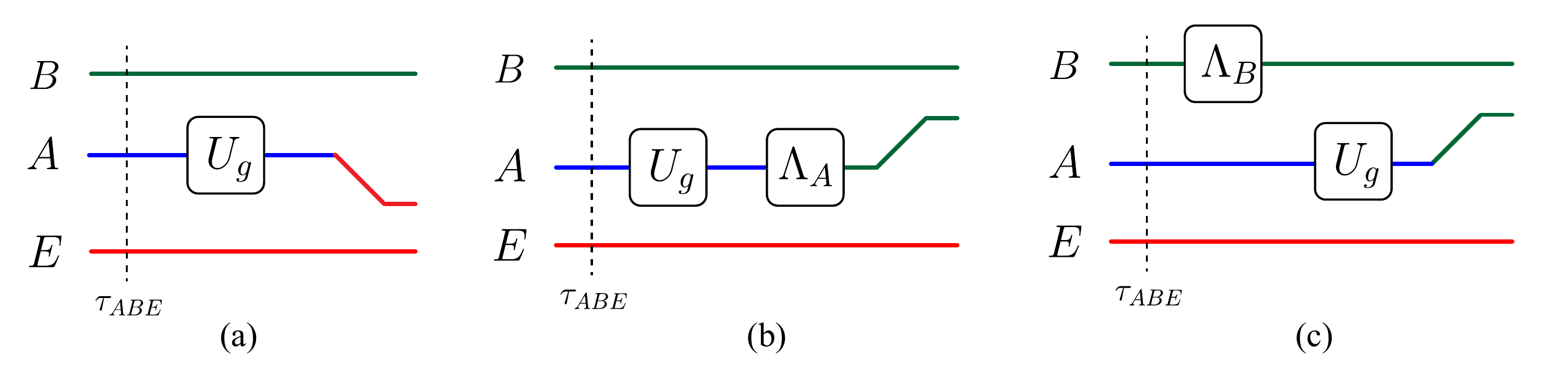}
	\caption{This figure illustrates the basic setting of private dense coding with general preshared state. 
	(a) In the worst case, the encoded state is intercepted by Eve.  (b) Without interception by Eve, Bob receives Alice’s state through channel $\Lambda_A$. (c) The case when there exists a channel $\Lambda_B$ such that $\Lambda_A(\tau_{AB})= \Lambda_B(\tau_{AB})$.}
\end{figure}

Our task is the secure transmission of the message 
$M \in \ca{M}$ from Alice to Bob by using $n$ copies of their shared state
under the following conditions;
Alice is allowed to freely communicate to Bob via a public channel. 
Alice can apply a unitary $U_g$ for $g \in G$
on her local states and sends it to Bob
after Bob  receives the quantum systems 
sent by Alice, he acknowledges the fact.
This task has the following criteria.

\begin{description}
\item[(C)] Completeness:
The PDC protocol is $\epsilon_C$-complete 
if the error verification aborts with probability no larger than $\epsilon_C$ 
when the state is not intercepted and the channel from Alice to Bob is exactly $\Lambda_A$.

 \item[(S)] Soundness: This part is composed of two conditions.
\begin{description}
\item[(S1)] Secrecy:
The PDC protocol is $\epsilon_E$-secure 
if the amount of the information leakage to Eve is upper bounded by $\epsilon_E$ 
even in the worst case, i.e., when Eve intercepts 
the whole transmitted quantum state via noiseless channel. 
The information leakage is evaluated by
\begin{align}
d(M ,\bar{E})_{\tau_{M\bar{E},{\rm f} }} 
:=
\min_{\sigma_{\bar{E}}}
\norm{
\tau_{M\bar{E},{\rm f}}- P_\ca{M} \otimes \sigma_{\bar{E},{\rm f}}   
}_{1},
\Label{CLP}
\end{align}
where
$\bar{E}$ expresses all the systems that Eve obtains during the protocol including 
the intercepted system $A$.
$\tau_{M\bar{E},{\rm f}}$ expresses the final state of our protocol,
and $P_\ca{M}$ is the uniform distribution for the message $M$.

\item[(S2)] Reliability:
The PDC protocol is $\epsilon_B$-reliable 
if the relation $M= \hat{M}$
holds with significance level $\epsilon_B$,
where $M$ is Alice's original message
and $ \hat{M}$ expresses Bob's recovered message.
Significance level is 
the probability of accepting the hypothesis to be shown, 
given that the hypothesis is assumed to be incorrect.
Now, $\Ab$ 
is the event that Bob aborts
the protocol so that $\Ab^c$ is the event that Bob does not abort the protocol.
Hence, $\Pr[\Ab^c|M=m,\hat{M}=\hat{m}]$ expresses the probability that 
Bob does not abort the protocol 
when
Alice's original message and Bob's recovered message
are $m$ and $\hat{m}$, respectively.
Then, the above condition is rewritten as
$\sup_{m\ne \hat{m} }\Pr[\Ab^c|M=m,\hat{M}=\hat{m}] \le \epsilon_B$.
\end{description}
\end{description}

For a given protocol $P_n$ for private dense coding with $n$ copies of $\tau_{ABE}^{\otimes n}$,
its parameters $\epsilon_C$, $\epsilon_E$, and $\epsilon_B$
are denoted by 
$\epsilon_C(P_n)$, $\epsilon_E(P_n)$, and $\epsilon_B(P_n)$, respectively.
Also, the rate of the message $\frac{\log |\ca{M}|}{n}$
is denoted by $R(P_n) $.
In this setting, 
Soundness is more important than completeness
because soundness shows the quality of our private communication when it is successful.
Hence, 
the parameter of completeness $\epsilon_C$ is not required to be too small.
For the parameter $\epsilon_C$ for Completeness, 
the value $1-\epsilon_C$ is called a power in the context of statistical hypothesis testing.
While there is no definite value for power, Cohen \cite{Cohen}
suggested $0.8$ as a conventional level, which corresponds to the choice $\epsilon_C=0.2$. 
However, a larger power (i.e., smaller $\epsilon_C$) can be chosen in practice. 
The reason is the following; 
Consider the case when Alice and Bob abort the protocol mistakenly while it runs without Eve's interception.
They can run the same protocol again until it is successful.
In contrast, the two parameters $\epsilon_E$ and $\epsilon_B$
of soundness need to be very small rigorously.
Hence, the two parameters $\epsilon_E$ and $\epsilon_B$
are called security parameters
because they express our confidence level.

\subsection{Protocol with finite-length setting}
Section \ref{S7-A} constructs our PDC protocol by combining 
wire-tap code and error verification.
Since the cost of error verification is negligible, 
it is sufficient to consider the cost for wire-tap code.
That is, we focus on
a cq wire-tap channel model \cite{Cai2004,Devetak2005,Devetak-Shor,Hayashi2015}, where
the channel $W_B$ to Bob is the cq-channel $g (\in G)\mapsto 
W_B(g):=\Lambda_A(U_g \tau_{AB} U_g^\dagger)=U_g \Lambda_A( \tau_{AB}) {U_g}^\dagger
$,
and 
the channel $W_E$ to Eve is the cq-channel $g (\in G)\mapsto W_E(g):=U_g \tau_{AE} U_g^\dagger$.

For $n$ copies of the state $\tau_{ABE}$,
combining wire-tap codes and error verification 
with the parameters $R_1$, $R_2$, and $\mathsf{t}$,
Section \ref{S7-A} constructs a PDC protocol $P_n$ of private dense coding
such that $R(P_n)= R_1-R_2-\frac{\mathsf{t}}{n}$ and
\begin{align}
\epsilon_C(P_n) &\le 4 \min_{0\le t \le 1} 
	2^{s n \left[ R_1 -\log d_A + H_{1-t}^\downarrow(A|B| \Lambda_A(\tau_{AB})) \right]} \Label{eq:B1}\\
	\epsilon_E(P_n)&\le  \min_{0\le t \le 1} 
	2^{-\frac{2s}{1+t}}2^{\frac{tn}{1+t}(- R_2 +\log d_A
	- \tilde{H}_{1+t}^\uparrow(A|E| \tau_{AE}))} \Label{eq:B2}\\
\epsilon_B(P_n) &\le  2^{-\mathsf{t}},\Label{eq:B3}
\end{align}
where $d_A$ is the dimension of $\ca{H}_A$.
Here, $R_1$ and $R_2$ are called the error correcting rate and the sacrificed rate for secrecy, respectively, and
$t$ is called the sacrificed length for reliability.
In the above formulas, two types of
\renyi conditional entropies $H_{1-t}^\downarrow(A|B| \Lambda_A(\tau_{AB}))$
and $\tilde{H}_{1+t}^\uparrow(A|E| \tau_{AE})$ are used.
\begin{align}
	H_{1+t}^\downarrow(A|B|\rho_{AB}) &\coloneqq - D_{1+t}(\rho_{AB}\|I_A \otimes \rho_B),\Label{eq:DefIG} \\
	\tilde{H}_{1+t}^\uparrow (A|B|\rho_{AB}) &\coloneqq - \inf_{\sigma_B\in \ca{D}(\ca{H}_B)} \tilde{D}_{1+t}(\rho_{AB}\|I_A \otimes \sigma_B) ,
\end{align}
where $I$ is the identity operator and 
we put the measured quantum state at the last entry of information quantities.
The Petz version \cite{Petz1986} and the sandwiched version \cite{MuellerLennert2013,Wilde2014} of quantum relative entropy
are defined as follows.
For $t \in (-1,\infty)$,  
\begin{align}
	D_{1+t}(\rho\|\sigma) &\coloneqq \frac{1}{t} \log \tr \rho^{1+t}\sigma^{-t}, \\
	\tilde{D}_{1+t}(\rho\|\sigma) &\coloneqq \frac{1}{t} \log \tr \sandwc{\rho}{\sigma}.
\end{align}
The sandwiched form is always no larger than the Petz form 
according to Araki-Lieb-Thirring trace inequality \cite{Araki1990}.
In the $t=0$ case, the above quantities are defined by taking the limit $t\rightarrow 0$, 
which is exactly $D(\rho\|\sigma)\coloneqq \tr \rho (\log \rho -\log \sigma)$. 
Hence, $\lim_{t\to 0}H_{1+t}^\downarrow(A|B|\rho_{AB}) =
\lim_{t\to 0}\tilde{H}_{1+t}^\uparrow (A|B|\rho_{AB})=
H(A|B|\rho):=H(AB|\rho)-H(B|\rho)$, where 
$H(AB|\rho)=-\tr \rho_{AB}\log \rho_{AB}$ 
and $H(B|\rho)=-\tr \rho_{B}\log \rho_{B}$. 

In addition, as explained in Section \ref{S5}, 
when the following conditions hold, it is possible to make 
a PDC protocol with small calculation complexity. 
\begin{description}
\item[(B1)]
The group $G$ forms a vector space over a finite field $\FF_q$.
\item[(B2)] The states $\{ U_g \Lambda_A( \tau_{AB}) U_g^\dagger\}_{g \in G}$
are commutative with each other.
\end{description}

\subsection{Asymptotic setting}
In the asymptotic setting of the PDC model, we impose the condition  
$\epsilon_C(P_n),\epsilon_E(P_n),\epsilon_B(P_n) \to 0$ as $n \to \infty$
for a sequence of PDC protocols $\{P_n\}$ from a theoretical viewpoint.
Under the above conditon, 
the rate $ \lim_{n\to \infty}R(P_n)$ is called an achievable private rate for 
a private dense coding setting $\text{PDC}(\tau_{ABE}, \{U_g\}_{g\in G},\Lambda_A)$. 
The supremum of achievable private rates is called 
the private capacity for $\text{PDC}(\tau_{ABE}, \{U_g\}_{g\in G},\Lambda_A)$,
and is written as $C(\tau_{ABE}, \{U_g\}_{g\in G},\Lambda_A)$.
As stated as Lemma \ref{lemma:AP},
the capacity of the above wire-tap channel model $(W_E,W_B)$
equals the private capacity for $\text{PDC}(\tau_{ABE}, \{U_g\}_{g\in G},\Lambda_A)$.
However, 
the capacity of wire-tap channel does not have a known single-letterized expression while the classical setting has \cite{Csiszar1978}.
Hence, we consider an achievable private rate $R_{*}$ 
by using the formulas \eqref{eq:B1}, \eqref{eq:B2}, and \eqref{eq:B3}.
For this aim, we define the map ${\cal G}$ as
${\cal G}(\rho):= \sum_{g \in G}\frac{1}{|G|}U_g \rho U_g^\dagger $.

Since $\lim_{t \to 0}\tilde{H}_{1+t}^\uparrow(A|E| \tau_{AE})$ equals
\begin{align}
R_{1,*}:=H( \Lambda_A \circ {\cal G} (\tau_{AB}))-
H( \Lambda_A (\tau_{AB}))
=H( {\cal G} (\Lambda_A (\tau_{AB})))-
H( \Lambda_A (\tau_{AB}))\Label{ED1},
\end{align}
the parameter $\epsilon_C(P_n)$ goes to zero with $R_1< R_{1,*}$.
Similarly, since $\lim_{t \to 0} \tilde{H}_{1+t}^\uparrow(A|E| \tau_{AE})$ equals
\begin{align}
R_{2,*}=H(  {\cal G} (\tau_{AE}))-H( \tau_{AE}),\Label{ED2}
\end{align}
the parameter $\epsilon_E(P_n)$ goes to zero with $R_2> R_{2,*}$.
When $t=\sqrt{n}$, $\epsilon_B(P_n)$ and $\frac{t}{n}$ go to zero.
Hence, the following is an achievable private rate;
\begin{align}
R_*=H( {\cal G} (\Lambda_A (\tau_{AB})))-
H( \Lambda_A (\tau_{AB}))-
H(  {\cal G} (\tau_{AE}))+H( \tau_{AE}) .\Label{E1}
\end{align}

In fact, the above achievable rate can be considered as the difference between two mutual informations.
When we denote the choice of an element $g \in G$ by the random variable $X$,
\eqref{ED1} and \eqref{ED2} become $R_{1,*}=I(X;BB'), R_{2,*}=I(X;AE)$ and
the achievable private rate of the above wire-tap channel model
is given as $I(X;BB')-I(X;AE)$\cite{Cai2004,Devetak2005,Devetak-Shor,Hayashi2015}.

\subsection{Asymptotic analysis with noiseless channel}
For further analysis, we assume that 
$\Lambda_A$ is the noiseless channel
and $\{U_g\}_{g\in G}$ is irreducible.
Then the rates $R_*,R_{1,*},$ and $R_{2,*}$ are simplified to
\begin{align}
R_{1,*}=\log d_A-H(A|B)_\tau, \quad
R_{2,*}=\log d_A-H(A|E)_\tau, \quad
R_*=H(A|E)_{\tau}-H(A|B)_\tau
\end{align}
because 
$H(  {\cal G} (\tau_{AB}))=H(  \frac{I_A}{d_A}\otimes \tau_{B}))
=\log d_A+ H( \tau_{B})$,
$H(  {\cal G} (\tau_{AE}))=H(  \frac{I_A}{d_A}\otimes \tau_{E}))
=\log d_A+ H( \tau_{E})$,
where $d_A:=\dim {\cal H}_A$.
Here, $ H(A|E)_\tau$ expresses the conditional entropy when the density matrix is $\tau_{AE}$.
When the total state $\tau_{ABE}$ is a pure state
and $\Lambda_A$ is the noiseless channel,
we have the relations
$H( \tau_{AE})=H( \tau_{B})$ and
$H( \tau_{AB})=H( \tau_{E})$.
Hence, the rate $R_*$ is simplified to
\begin{align}
R_*=2 (H(\tau_{A,E})-H(\tau_E))
=2 H(A|E)_\tau=-2 H(A|B)_\tau.
\Label{E1T}
\end{align}

In fact, as shown in Appendix \ref{A1}, when $\Lambda_A$ is the noiseless channel, $\tau_{ABE}$ is a pure state, and $\tau_{AB}$ is maximally correlated, i.e., 
there exist bases $|v_j^A\rangle$ and $|v_j^B\rangle$ on
$\mathcal{H}_A$ and $\mathcal{H}_B$ such that
$\tau_{AB}=\sum_{j,j'}a_{j,j'}|v_j^A,v_j^B\rangle
\langle v_{j'}^A,v_{j'}^B|$,
then the above cq wire-tap channel $W_B, W_E$ is degraded \cite{Devetak-Shor,hayashi2016quantum}, i.e., 
there exists a TP-CP map $\Gamma$ such that
$W_E(g)=\Gamma(W_B(g))$.
In this case, as shown in Corollary \ref{A72}, due to the group symmetric condition,
 the quantity \eqref{E1T} gives the private capacity of our PDC model, i.e., the maximum secure transmission rate.

Here, we should remark the relation with 
the method studied in the preceding research \cite{Devetak2005a}.
The reference \cite{Devetak2005a} considers
the secure key distillation from a preshared state $\tau_{ABE}$ with the same condition for this subsection.
They showed that their one-way method achieves 
the key generation rate $H(A|E)_\tau$.
However, since 
our method is allowed to use quantum communication from Alice to Bob,
it achieves twice of their rate.

\subsection{Reduction to noiseless case}
In the latter part of the above subsection, we assume that the channel $\Lambda_A$
is noiseless.
When our model with noisy channel $\Lambda_A$
satisfies the following condition,
its analysis can be reduced to the case with the noiseless channel.
\begin{description}
\item[(A1)]
There exists a TP-CP map $\Lambda_B$ on $\mathcal{H}_B$
such that $\Lambda_A( \tau_{AB})=\Lambda_B( \tau_{AB})$.
\end{description}
Indeed, 
the reliability of $\text{PDC}(\tau_{ABE},\{U_g\}, \Lambda_A)$
is equivalent to 
the reliability of $\text{PDC}(\Lambda_B(\tau_{ABE}),\{U_g\}, \id_A)$ because
the reduced state of $\Lambda_B(\tau_{ABE})$ to 
$\mathcal{H}_{A}\otimes \mathcal{H}_B$
equals $\Lambda_A( \tau_{AB})$,
where $\id_A$ is the identical channel.
The secrecy is also equivalent
because the reduced state of $\tau_{ABE}$ to 
$\mathcal{H}_{A}\otimes \mathcal{H}_E$
equals 
the reduced state of $\Lambda_B(\tau_{ABE})$ to 
$\mathcal{H}_{A}\otimes \mathcal{H}_E$.
Therefore, 
the analysis with preshared state $\tau_{ABE}$ and noisy channel $\Lambda_A$
is reduced to 
the analysis with preshared state $\Lambda_B(\tau_{ABE})$ and the noiseless channel.

\section{Application to Weyl-Heisenberg representation \Label{SWH}}
\subsection{PDC model with Weyl-Heisenberg representation}
In this section, we discuss the private dense coding with preshared state
with Weyl-Heisenberg representation. That is,
the dimension of space $\ca{H}_A$ is a prime $d_A=p$,
the group $G$ is Weyl-Heisenberg group $\mathbb{F}_p^2$ 
and
the group representation is given as a set of implementable operations as follows.
In addition, the rate $R_*$ can be achieved
by a protocol with small calculation complexity 
as explained latter.
The $\mbf{X}$ and $\mbf{Z}$ operators on the $p$-dimensional quantum system are defined as
\begin{align*}
	&\mbf{W}(x,z) = \mbf{X}^x \mbf{Z}^z\\
	&\mbf{X} = \sum_{j\in\mathbb{F}_p} \ket{j+1}\bra{j},\\
	&\mbf{Z} = \sum_{j\in\mathbb{F}_p} \omega^j\ket{j}\bra{j},\end{align*}
with $\omega = e^{i2\pi/p}$.
In this case, 
the private dense coding with preshared state is the same as
the protocol discussed in \cite{Wang2005,Wu2019}.

As a typical noise model, we employ a generalized Pauli channel acting on 
a $d$-dimensional quantum system defined as 
\begin{align}
	\Lambda[P_{XZ}](\rho) = \sum_{(x,z)\in \mathbb{F}_p^2} P_{XZ}(x,z) \mbf{W}(x,z) \rho \mbf{W}(x,z)^\dagger.\Label{COP}
\end{align}
That is, we assume that the preshared state is generated by 
transmission of a maximally entangled state $\ket{\Phi} = \frac{1}{\sqrt{p}} \sum_i \ket{i}_A\ket{i}_B$ 
from Bob to Alice via the channel 
$\Lambda[P_{XZ}]$ acting on $\ca{H}_A$.
In this case, the joint state $\tau_{AB}$ on ${\cal H}_A\otimes {\cal H}_B$ is given as the generalized  Bell diagonal state;
\begin{align}
\rho[P_{XZ}]:=  \sum_{(x,z)\in \mathbb{F}_p^2} P_{XZ}(x,z) \mbf{W}(x,z) |\Phi\rangle \langle \Phi| \mbf{W}(x,z)^\dagger.
\Label{BOR}
\end{align}

\begin{figure}[!t] 
	\centering
	\includegraphics[width=6in]{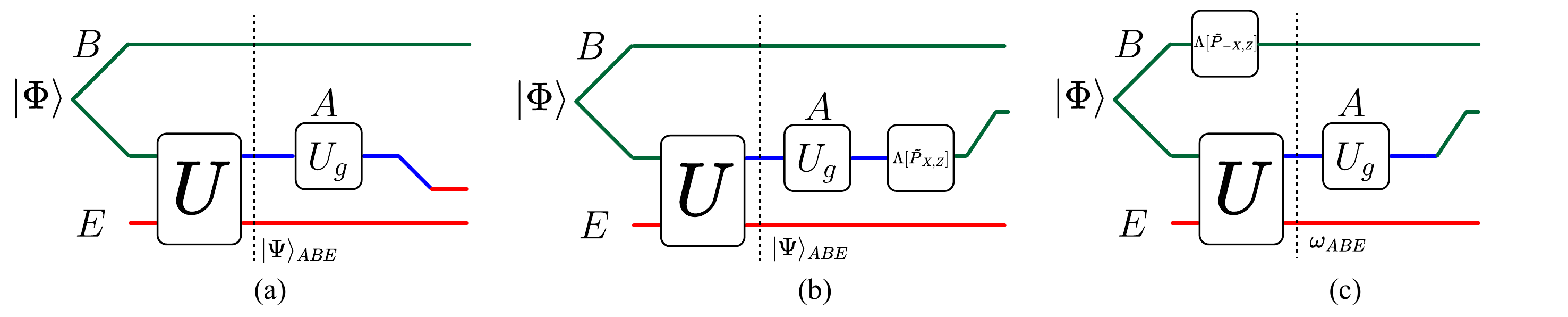}
	\caption{This figure illustrates the setting $\text{PDC}(\ket{\Psi},\{\mbf{W}(x,z)\}, \Lambda[\tilde{P}_{XZ}]_{A})$. 
	$U$ is the unitary that reduces to the Pauli channel $\Lambda[P_{XZ}]$ by tracing out system $E$.
	Entanglement distribution is done via the application of the unitary $U$ to the 
	maximally entangled state. 
	$\ket{\Psi}_{ABE}$ is the output state of the above entanglement distribution.
	(a) In the worst case, Alice's state is intercepted by Eve through a noiseless channel. 
	(b) Without interception by Eve, Bob receives Alice's state through channel $\Lambda[\tilde{P}_{X,Z}]$.
	(c) The noiseless reduction of original setting.} \label{fig:WH}
\end{figure}

To guarantee the secrecy under the worst case,   
we assume that Eve controls all the environment of the channel $\Lambda[P_{XZ}]$.
Hence, the state on the system ${\cal H}_A\otimes {\cal H}_B\otimes {\cal H}_E$
is the pure state $\tau_{ABE}= \ket{\Psi}\bra{\Psi}$, 
i.e., the purification of $\rho[P_{XZ}]$,
which is given as
\begin{align}
 \ket{\Psi}_{ABE} &= \frac{1}{\sqrt{d}}\sum_{x,z}
\sqrt{P(x,z)} \mbf{W}(x,z)\ket{\Phi}_{AB} \ket{x,z}_E 
\Label{ACO}
\end{align}
Also, the channel $\Lambda_A$ from Alice to Bob is given as
another generalized Pauli channel
$\Lambda_{A}[\tilde{P}_{XZ}]$.
That is, we focus on private dense coding 
$\text{PDC}(\ket{\Psi},\{\mbf{W}(x,z)\}, \Lambda[\tilde{P}_{XZ}]_{A})$ as depicted in Fig. \ref{fig:WH}, 
where $(x,z)\in \mathbb{F}_p^{2}$ by default.
Since we have
\begin{align}
 \Lambda[\tilde{P}_{X,Z}]_A 
\circ \Lambda[{P}_{XZ}]_A(|\Phi\rangle \langle \Phi|)
= \Lambda[\tilde{P}_{-X,Z}]_B 
\circ \Lambda[{P}_{XZ}]_A(|\Phi\rangle \langle \Phi|),\Label{XPA}
\end{align}
we can alternatively apply the model 
$\text{PDC}(\omega_{ABE},\{\mbf{W}(x,z)\}, \id_A)$, 
where $\omega_{ABE} := \Lambda[\tilde{P}_{-X,Z}]_B (\ket{\Psi}\bra{\Psi})$.

In the private dense coding model $\text{PDC}(\omega_{ABE},\{\mbf{W}(x,z)\}, \id_A)$, it can be shown that 
the state received by Bob is always Bell diagonal (see Appendix \ref{app:Bell}), 
i.e., the condition (B2) holds.
Therefore, Bob can extract the information via 
the measurement $\Pi_{x,z} =\text{Proj}(\mbf{W}(x,z)\ket{\Phi})$ 
without any state demolition.
Once this measurement is applied, 
the channel from Alice to Bob is given as a classical channel
$W^c$, which is given as $W^c(x,z|x',z'):= (\tilde{P}_{XZ}*P_{XZ})
(x-x',z-z')$, where the convolution 
$\tilde{P}_{XZ}*{P}_{XZ}$ is defined by 
\begin{align}
	\tilde{P}_{XZ}*{P}_{XZ}(x,z) &:= \sum_{x'z'}
	\tilde{P}_{XZ}(x',z') {P}_{XZ}(x-x',z-z').
\end{align}
That is, Bob can apply classical decoding to the channel $W^c$
without any information loss.

Since the group $G$ forms a vector space over a finite field $ \mathbb{F}_p$
and the states received by Bob are commutative with each other in the above way,
the conditions (B1) and (B2) are satisfied so that
the rate $R_*$ can be achieved by a protocol with small calculation complexity. 
For this model, we have the following lemma, whose proof is shown in Appendix \ref{P-lemma:PI1}.
\begin{lemma} \Label{lemma:PI1}
The key information quantities in the setting $\text{PDC}(\omega_{ABE},\{\mbf{W}(x,z)\}, \id_A)$
can be expressed as follows.
\begin{align}
	H(A|B)_\omega
	&= H(X Z| \tilde{P}_{XZ}*{P}_{XZ}) - \log d_A \Label{eq:IY}\\
	H(A|E)_\omega &= \log d_A - H(XZ|P_{XZ}) \Label{eq:I1}\\
	\tilde{H}_{1-t}^\downarrow(A|B)_\omega &= H_{1-t} (\tilde{P}_{XZ}*{P}_{XZ}) -\log d_A  \Label{eq:HAB}\\
	\tilde{H}_{1+t}^\uparrow (A|E)_\omega &\ge \log d_A - H_{\frac{1}{1+t}}(P_{XZ}) \Label{eq:HAE},
\end{align}
where $H_{1+t}(\rho) := -D_{1+t}(\rho\|I)$.
\end{lemma}
By applying \eqref{eq:IY} and \eqref{eq:I1} to the formulas \eqref{ED1}, \eqref{ED2} and \eqref{E1}, 
the rates $R_*,R_{1,*},$ and $R_{2,*}$ are calculated as
\begin{align}
R_{1,*}=&2 \log d_A- H(\tilde{P}_{XZ}*{P}_{XZ})\Label{BXP}
\\
R_{2,*}=&H(P_{XZ})
\\
R_*=& 2\log d_A - H(P_{XZ})- H(\tilde{P}_{XZ}*{P}_{XZ}) .\Label{BXP3}
\end{align}

\subsection{Application of PDC protocol}\Label{SWH2}
The high-dimensional quantum secure direct communication (QSDC) \cite{Wang2005}
is similar to our PDC model. 
As its generalization,
we propose the following protocol based on our PDC model. 
The aim of the following protocol is 
that Alice transmits her message $M \in \ca{M}:=\FF_p^{n_2}$
to Bob by using two quantum channels, i.e.,
$\Lambda[P_{XZ}]$ from Bob to Alice and 
$\Lambda[\tilde{P}_{XZ}]$ from Alice to Bob,
and a free public channel in both directions.
In the following scenario, 
Bob distributes quantum states via $\Lambda[P_{XZ}]$ initially.
Next, Alice sends back the received system to Bob via
$\Lambda[\tilde{P}_{XZ}]$ after her coding operation.
While we assume there is no possibility that
the channel $\Lambda[P_{XZ}]$ from Bob to Alice is changed or intercepted,
Eve is assumed to receive the environment of the channel $\Lambda[P_{XZ}]$.
That is, by using a unitary $U$ from $\ca{H}_B$ to $\ca{H}_A\otimes \ca{H}_E$
whose reduction to $\ca{H}_A$ is $\Lambda[P_{XZ}]$,
 the transmission process from Bob to Alice is regarded as the application of 
the unitary $U$, and Eve has the system $\ca{H}_E$ of the output of the unitary $U$.
In addition, we consider that Eve intercepts the second quantum communication
from Alice to Bob in the worst case.
Due to the above assumption, Completeness (C) and Soundness (S) can be applied to 
this problem setting in the same way as the PDC model.

To state our protocol, 
we prepare 
a classical error correcting code $\varphi=(\varphi_e,\varphi_d)$ for 
$n$ uses of the classical channel $W^c$ with decoding error probability $\epsilon(\varphi)$, 
where $\varphi_e$ is a classical linear encoder that maps
$\FF_p^{n_1}$ to a linear subspace $\ca{L}$ of $\FF_p^{2n}$, and 
$\varphi_d$ is a classical decoder that maps  
$\FF_p^{2n}$ to $\FF_p^{n_1}$.
Here, $n_1$ is called the coding length of the code $\varphi$. 
Then, we prepare the message set $\ca{M}:=\FF_p^{n_2}$,
the set for covering variable $\ca{Y}:=\FF_p^{n_3}$, 
which will be used to cover the message in error verification,
the message set for wire-tap code
$\ca{M}':=\ca{Y}\times \ca{M}=\FF_q^{n_2+n_3}$, and
two sets of random seeds $\ca{S}:=\FF_p^{n_1-1}$,
 $\ca{S}':=\FF_p^{n_2+ n_3-1}$.
We prepare two universal hash function (UHF) families
$ f_S: \FF_p^{n_1} \to \ca{M}'$ and 
$ g_{S'}: \ca{M}'\to \ca{Y}$, which is defined in \eqref{E64} and \eqref{E65}, respectively.
Then, combining the encoder of classical error correcting code $\varphi_e$ and 
the random seed $S$, we define the function $\psi_S$ by \eqref{E66}.
Under the above preparation, we propose Protocol 1, 
which is also illustrated in Fig.\ref{fig:protocol}.

{\bf Protocol 1} 
\Label{protocol1}
\begin{algorithmic}
\STATE {\bf Entanglement distribution:}\quad 
Bob starts the protocol by generating $n$ entangled state $\ket{\Phi} = \frac{1}{\sqrt{d}}\sum_j\ket{j}_B\ket{j}_{A}$. 
Then he sends the system $A$ of the states to Alice and keeps the rest $B$ system.

\STATE {\bf Encoding:}\quad 
When Alice intends to send the message $M\in\ca{M}$, 
Alice generates random variables
$S\in \ca{S}$,
$S'\in \ca{S'}$,
$Y\in \ca{Y}$, and 
$L_2\in \FF_p^{n_1-(n_2+n_3)}$
independently according to the uniform distribution.
Alice chooses elements $X^n=(X_1X_2\ldots X_n):=\psi_S(Y,M,L_2)\in \FF_p^{2n}$.
Alice applies private dense coding operation $\mbf{W}(x_i,z_i)$ on the $i$-th state and sends them back to Bob.

\STATE {\bf Reception:}\quad 
Bob applies projective measurement $\Pi=\{\Pi_{x,z}\}$ with 
	$\Pi_{x,z} =\text{Proj}(\mbf{W}(x,z)_A\ket{\Phi})$ on the composite system of the received states and the kept states to obtain classical string 
	$\hat{X}^n=(\hat{X}_1\hat{X}_2...\hat{X}_n)$ with $\hat{X}_i\in\mathbb{F}_p^{2}$,
	where the projection operator on state $\ket{\psi}$ is denoted by $\text{Proj}(\ket{\psi})$.
	Specifically, if $p=2$, the measurement becomes Bell state measurement. Finally, 
	Bob acknowledges this reception to Alice via the public channel.

\STATE {\bf Public communication from Alice to Bob:}\quad 
Alice sends the variables $S$, $S'$, and $C:=g_{S'}(M,Y)$ to Bob via
the public channel.

\STATE {\bf Decoding \& Verification :}\quad 
Bob performs classical decoding to $\hat{X}^n$, i.e., he obtains
$(\hat{M},\hat{Y}):=f_{S}(\varphi_d(\hat{X}^n))$.
If $g_{S'}(\hat{M},\hat{Y})=C$, he accepts the message $\hat{M}$.
Otherwise, he aborts the protocol.
\end{algorithmic}

\begin{figure}[!t]
	\centering 
	\includegraphics[width=6in]{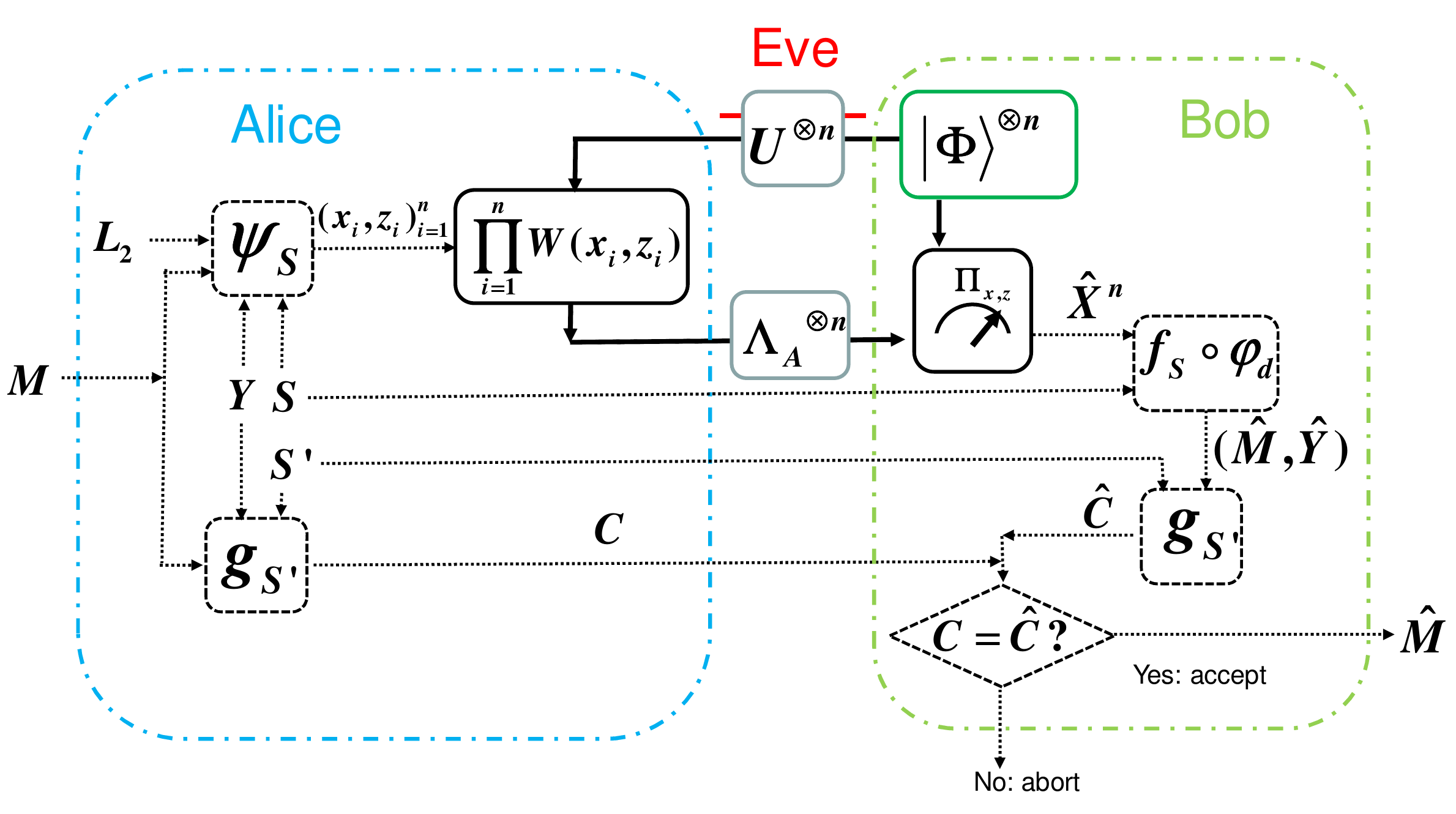}
	\caption{This figure illustrates Protocol 1.
	Real lines express quantum information flow.
	Dotted lines express classical information flow.
	$\ket{\Phi} = \frac{1}{\sqrt{d}}\sum_{i} \ket{i}_{A'}\ket{i}_B$ 
	is the initial state generated by Bob
	and $U$ is a unitary performed by Eve.
	In Alice's encoding part, $\psi_S$ is the encoder acting on the message $M$ and random number $Y,L_2$.
	$\mbf{W}(x_i,z_i)$ is the encoding operation on the $i$-th state.
	In Bob's decoding part, $\Pi_{x,z}$ is a generalized Bell state measurement and 
	the classical decoder $f_S \circ \varphi_d$ is a concatenation of the decoder of a classical error correcting code and a universal hash function.
	In addition, they decide if the protocol aborts by comparing the results
	of universal hash function $g_{S'}$ .
	Note that $C$ and the random seeds $S, S'$ are transmitted via a public channel.
	The explicit definitions for $\psi_S, g_{S'}, f_S$ are given in Section \ref{S7-A}.
	\Label{fig:protocol}}
\end{figure}

Due to \eqref{eq:errorA},
we find that this protocol has $\epsilon(\varphi)$-completeness.
Using \eqref{eq:TrdA} and \eqref{eq:TrdY} with \eqref{eq:HAB}, 
we find that 
this protocol has Soundness with parameters $\epsilon_E$ and $\epsilon_B$
when $n_1-(n_2+n_3)$ and $n_3$ are chosen to be 
$\hat{m}_2(\epsilon_E)$ and 
$\hat{m}_3(\epsilon_B)$ defined as
\begin{align}
\hat{m}_2(\epsilon_E) &:= \frac{1}{\log p}
\min_{0\le t\le 1} H_{\frac{1}{1+t}}(P_{XZ}) - \frac{1+t}{t }\log \epsilon_E - 2 \Label{ANO}\\
\hat{m}_3(\epsilon_B)&:= - \frac{\log \epsilon_B}{\log p}.
\Label{ANO2}
\end{align}
As explained in Proposition \ref{PRO1}, there exists an error correcting code $\varphi$
with the average decoding error probability $\epsilon_C$ and
coding length $\hat{m}_1(\epsilon_C)$
\begin{align}
\hat{m}_1(\epsilon_C) &= \frac{1}{\log p}
\max_{0\le t \le 1} 2\log d_A - H_{1-t} (P_{XZ}*P_{XZ}) + 
\frac{1}{t}(\log {\epsilon_C} - 2).
\Label{ANO1}
\end{align}

These relations show the existence of a protocol that is $\epsilon_C$-complete, 
$\epsilon_E$-secure, and $\epsilon_B$-reliable.
A simulation of the rates $R_1=\frac{\hat{m}_1(\epsilon_C)}{n},
R_2=\frac{\hat{m}_2(\epsilon_E)}{n},R_3=\frac{\hat{m}_3(\epsilon_B)}{n}$ and $R= R_1-R_2-R_3$ 
in qubit case is shown in Fig. \ref{fig:rates}.
The values defined in \eqref{ANO} and \eqref{ANO2}
have a practical meaning
because 
$\epsilon_E$-secure and $\epsilon_B$-reliable conditions can be satisfied
with these numbers $\hat{m}_2(\epsilon_E)$ and $\hat{m}_3(\epsilon_B)$
and calculation complexity $O(n\log n)$, as explained in Section \ref{S7-A}.
However, \eqref{ANO1} does not have such a practical meaning for completeness
because it assumes the combination of the random coding and maximum likelihood decoder.
In contrast, the asymptotic rate \eqref{BXP} has a practical meaning 
because there exist error correcting codes to achieve the rate \eqref{BXP} with a small calculation complexity
as explained in Section \ref{S7}
while their finite-length analysis is not so simple.
 
\begin{figure}[!t]
	\centering 
	\includegraphics[width=3in]{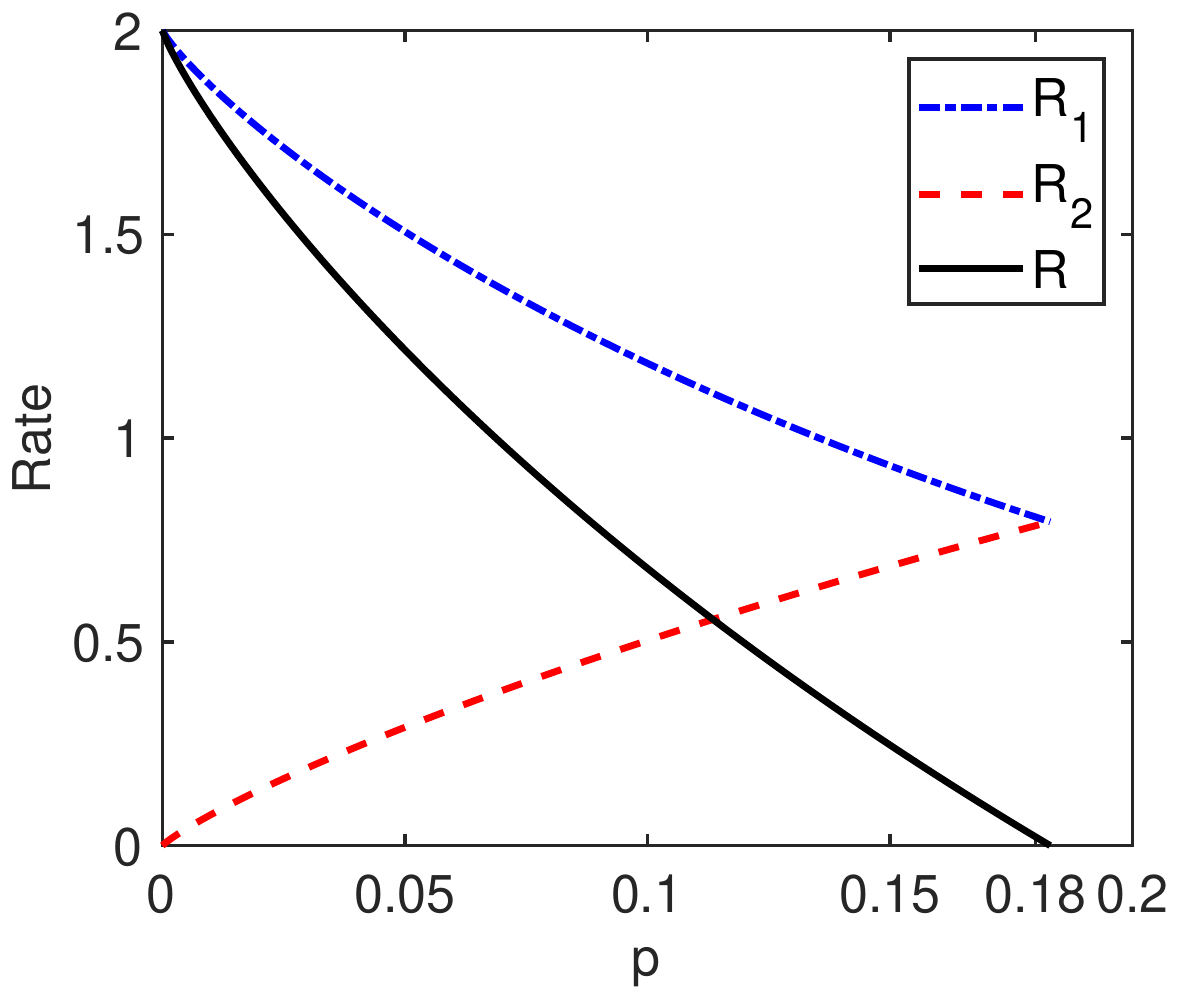}
	\includegraphics[width=3in]{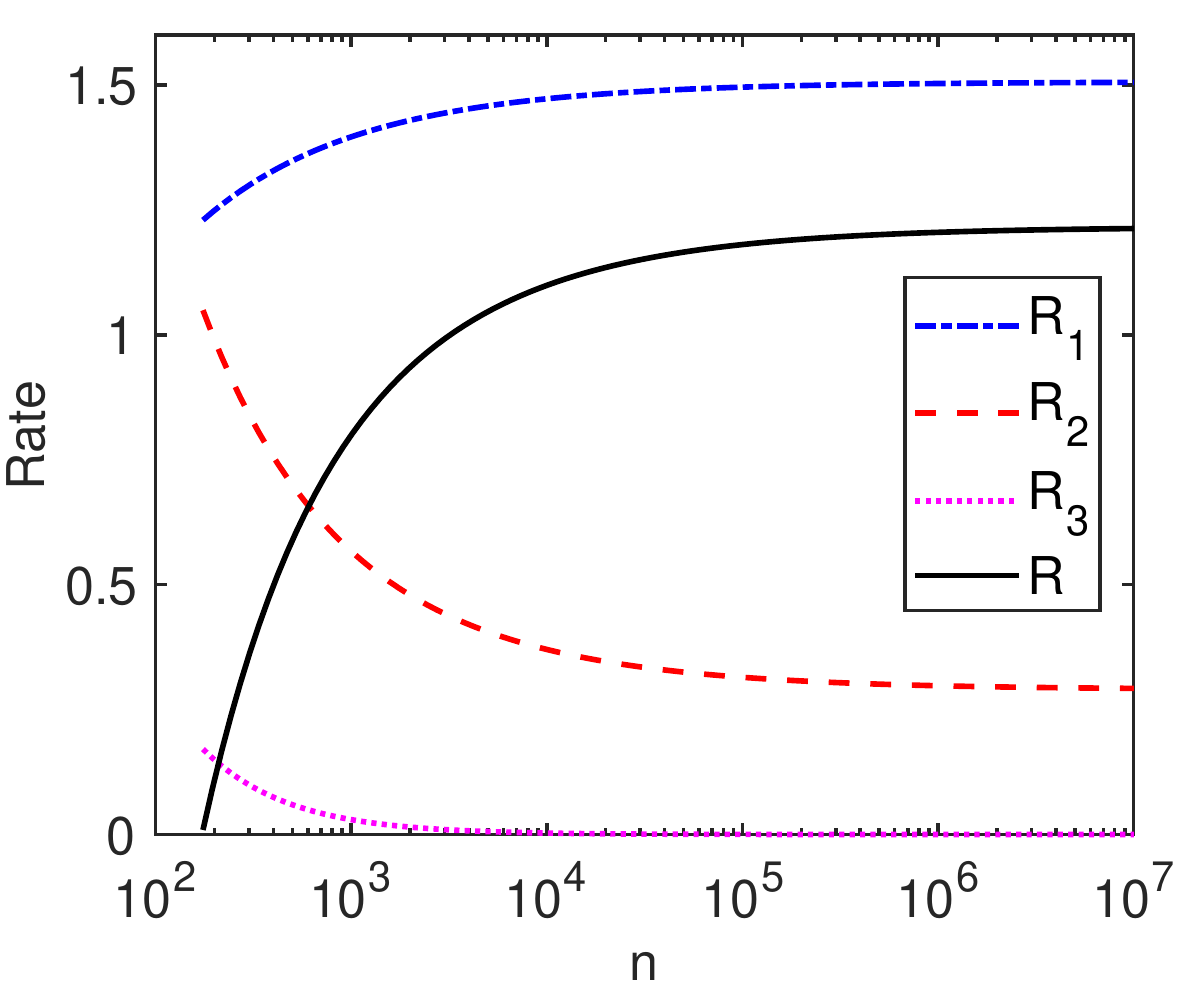}
	\caption{ This figure illustrates the error correcting rate $R_1$, the sacrificed rate for secrecy $R_2$,
	the sacrificed rate for error verification $R_3$ and the message rate $R= R_1 -R_2 -R_3$ 
	in asymptotic and finite block length settings.
	(a) Asymptotic rate in depolarizing channel $\ca{E}(\rho) = (1-p)\rho + p \rho_{\mix}$. 
	Note that $R_3=0$ asymptotically and the message rate reaches zero when $p\approx 0.18$.
	(b) Finite length rate in depolarizing channel with depolarizing probability $p=0.05$ 
	with criteria $\epsilon_C \le 0.2, \epsilon_B \le 10^{-9},\epsilon_E \le 10^{-9} $. 
	It is shown that the rates almost achieve the asymptotic limit at block length $10^{6}$. \Label{fig:rates}}
\end{figure}

\section{Application to unknown state with Weyl-Heisenberg representation}\Label{S4}
In the PDC model, we consider the case when the preshared state $\tau_{ABE}$ is unknown. This case corresponds to the case when the channel from Bob to Alice is unknown in the model stated in Subsection \ref{SWH2}.
In this case, Alice and Bob need to estimate it before their secure communication.
As a simple problem setting, we assume that 
Alice and Bob share $n'$ copies of the state $\tau_{AB}$
while they do not know the form of $\tau_{AB}$,
which is called the independent and identical density (i.i.d.) condition.
In the worst case,
Eve controls the whole of the environment system of the state 
$\tau_{AB}^{\otimes n'}$.
To satisfy the secrecy requirement, 
Alice and Bob need to identify the form of $\tau_{AB}$
and assume the worst case.

Suppose that Alice and Bob use $n'-n$ copies for the estimation of 
$\tau_{AB}$.
In this case, they estimate 
$\tau_{AB}$ by using only two-way LOCC, which is often called local tomography.
However, the estimation of the unknown state without any assumption
requires many types of measurements.
If $\mathcal{H}_A$ and $\mathcal{H}_B$ have the same dimension and 
the unknown state $\tau_{AB}$ is a Bell diagonal state $\rho[P_{XZ}]$,
we can reduce the number of measurements for the estimation
as explained in Appendix \ref{MKA}.

When $\mathcal{H}_B$ is the same-dimensional as 
$\mathcal{H}_A$ and $\tau_{AB}$ is a general state,
we can apply discrete twirling to the state $\tau_{AB}$:
\begin{align}
T(\tau_{AB}):=
\frac{1}{d^2}\sum_{x,z} 
(\mbf{W}(x,z)_A \otimes \overline{\mbf{W}(x,z)}_B) \tau_{AB}
(\mbf{W}(x,z)_A \otimes \overline{\mbf{W}(x,z)}_B)^\dagger,
\Label{BAP}
\end{align}
where $\overline{\mbf{W}(x,z)}_B$ is the complex conjugate of 
$\mbf{W}(x,z)_B$.
It is known that the above state is a
Bell diagonal state \cite[Example 4.22]{hayashi2017group}, \cite{PhysRevA.68.012301,PhysRevA.54.3824}.
Hence, when Alice and Bob apply the above discrete twirling 
and apply the estimation of a Bell diagonal state,
they can apply private dense coding whose 
shared state is the estimated Bell diagonal state.
However, as shown in Appendix \ref{MKA}, 
the twirled state $T(\tau_{AB})$ can be estimated by applying the estimation method given in Appendix \ref{MKA} to the state $\tau_{AB}$.

Therefore, combining the above method with Protocol 1, 
we propose Protocol 2,  
in which the encoding and decoding processes
contain the procedure for the discrete twirling.
In the following protocol, we assume that 
the communication channel from Bob to Alice is given as the $n'$ times use of the same 
channel from Bob to Alice while the channel is not known.

	{\bf Protocol 2}
	\begin{algorithmic}
	\STATE {\bf Entanglement distribution:}\quad 
	Bob starts the protocol by generating $n'$ entangled state $\ket{\Phi} = \frac{1}{\sqrt{d}}\sum_j\ket{j}_B\ket{j}_{A'}$. Then he sends the $A'$ halves of the states to Alice and keeps the rest system $\ca{H}_B$. Then, they share state $\tau_{AB}^{\otimes n'}$.
	\STATE {\bf Estimation:}\quad 
	Alice and Bob choose $n'-n$ samples,
	and estimate the twirled state $T(\tau_{AB})$ of
	the shared state $\tau_{AB}$ by using the method explained in Appendix \ref{MKA}. 
	Based on the estimation of $\tau_{AB}$, they decide their error correcting code $\varphi$ and the integers $n_1,n_2,n_3$.
	\STATE {\bf Encoding:}\quad 
	Alice generates
	a string $\bar{X}^n=(\bar{X}_1,\bar{X}_2, \ldots,\bar{X}_n)$ 
	with $X_i = (x_i,z_i)\in\mathbb{F}_d^{2}$ according to the uniform distribution. 
	Alice encodes her message $M$ into a string $X^n=(X_1X_2...X_n)
	:=\psi_S(Y,M,L_2)$ with $X_i\in\mathbb{F}_p^{2}$.
	For every elements $(x_i,z_i)\coloneqq \bar{X}_i+X_i$, 
	Alice applies operation $\mbf{W}(\bar{x}_i,\bar{z}_i) \mbf{W}(x_i,z_i)$ on the remaining $i$-th state and sends them back to Bob.
	\STATE {\bf Pubic communication from Alice to Bob 1:}\quad 
	Alice sends the string $\bar{X}^n$ to Bob via the public channel.
	\STATE {\bf Reception:}\quad 
		Bob applies unitaries $\overline{\mbf{W}(\bar{x}_1,\bar{z}_1)}\cdots
	\overline{\mbf{W}(\bar{x}_n,\bar{z}_n)} $ to 
	$\mathcal{H}_B^{\otimes n}$, where
	$(\bar{x}_i,\bar{z}_i):= \bar{X}_i$.
Then, 
Bob applies projective measurement $\Pi=\{\Pi_{x,z}\}$ 
	on the composite system of the received states and the kept states to obtain classical string 
	$\hat{X}^n=(\hat{X}_1\hat{X}_2...\hat{X}_n)$. 
	\STATE {\bf Pubic communication from Alice to Bob 2:}\quad 
      The same procedure as the same step of Protocol 1.
\STATE {\bf Decoding \& Verification :}\quad 
Bob performs classical decoding to $\hat{X}^n $, i.e., he obtains
$(\hat{M},\hat{Y}):=f_{S}(\varphi_d(\hat{X}^n))$.
If $g_{S'}(\hat{M},\hat{Y})=C$, he accepts the message $\hat{M}$.
Otherwise, he aborts the protocol.
	\end{algorithmic}

Since the estimation method given in Appendix \ref{MKA} works well,
due to the discussion in Subsection \ref{SWH2},
this protocol achieves the rate given in \eqref{BXP3}
with the limit $n \to \infty$.
For further analysis on the above protocol, notice the relation
\begin{align}
&W(\hat{x},\hat{z})_A \otimes (\overline{W(\bar{x},\bar{z})}_B)^\dagger|\Phi\rangle 
\langle \Phi| W(\hat{x},\hat{z})^\dagger_A\otimes \overline{W(\bar{x},\bar{z})}_B
\Label{KM1}\\
=&
W(\hat{x}+\bar{x},\hat{z}+\bar{z})_A \otimes I_B|\Phi\rangle 
\langle \Phi| W(\hat{x}+\bar{x},\hat{z}+\bar{z})^\dagger_A\otimes I_B.
\Label{KM2}
\end{align}
The operator \eqref{KM1} expresses a positive operator-valued measurement (POVM) element of the unitary $\overline{\mbf{W}(\bar{x}_i,\bar{z}_i)}$ and the measurement $\Pi$ as a whole performed in Reception step. 
The above relation shows that Bob can apply the measurement with POVM elements
given in \eqref{KM2} instead of \eqref{KM1}.
That is, when the outcome of the measurement corresponding to \eqref{KM2} by 
$(\underline{x},\underline{z}):= (\hat{x}+\bar{x},\hat{z}+\bar{z})$, 
Bob outcome $(\hat{x},\hat{z})$ of the above protocol is given as
$(\underline{x}-\bar{x},\underline{z}-\bar{z})$.
Therefore, Protocol 2 is converted to the following protocol.

	{\bf Protocol 3}
	\begin{algorithmic}
	\STATE {\bf Same steps as Protocol 2:}\quad 
	They make the same steps as Entanglement distribution, Estimation, and Encoding as Protocol 2.
	\STATE {\bf Reception:}\quad 
	Bob applies projective measurement $\Pi=\{\Pi_{x,z}\}$ 
	on the composite system of the received states and the kept states to obtain classical string 
	$\underline{X}^n=(\underline{X}_1\underline{X}_2...\underline{X}_n)$.
\STATE {\bf Pubic communication from Alice to Bob:}\quad 
Alice sends the variables $S$, $S'$, $C:=g_{S'}(M,Y)$, and $\bar{X}^n$ to Bob via
the public channel.
\STATE {\bf Decoding \& Verification :}\quad 
Bob performs classical decoding to $\underline{X}^n  - \bar{X}^n$, i.e., he obtains
$(\hat{M},\hat{Y}):=f_{S}(\varphi_d(\underline{X}^n- \bar{X}^n))$.
If $g_{S'}(\hat{M},\hat{Y})=C$, he accepts the message $\hat{M}$.
Otherwise, he aborts the protocol.
	\end{algorithmic}

\section{Modular code for quantum wire-tap channel}\Label{S5}
\subsection{Formulation for quantum wire-tap channel}
This section proposes a modular coding scheme with fixed error correcting code for quantum wire-tap channels and analyze its performance in finite length setting
because this code construction is used for our PDC protocol.
In the wire-tap channel model, the legitimate user Alice wants to send messages to another legitimate user Bob through a channel reliably and secretly in the presence of an eavesdropper Eve. The classical wire-tap channel has been extensively investigated since the debut of Wyner's wire-tap channel model \cite{Wyner1975}, which was subsequently generalized by Csisz\'{a}r and K\"{o}rner \cite{Csiszar1978}. The quantum wire-tap channel was also explored both in asymptotic setting \cite{Devetak2005,Cai2004} and finite setting \cite{Hayashi2015}. It acts as an approach to analyze the secrecy of QKD protocols \cite{Devetak2005a} and QSDC protocols \cite{Qi2019,Wu2019}.

Here we focus on classical-quantum (cq) wire-tap channel, which consists of two cq channels: $W_B: x(\in \ca{X})\rightarrow W_{B}(x)$ from Alice to Bob and 
$W_E: x(\in \ca{X})\rightarrow W_{E}(x)$ from Alice to Eve. 
Here, $W_{B}(x)$ and $W_{E}(x)$ are states on quantum systems
$\ca{K}_B$ and $\ca{K}_E$, respectively.
Alice's and Bob's procedures
are called an encoder $\Gamma$ 
and a decoder $\Pi$, respectively.
When they use $n$ times of the above channel,
an encoder is a map $\Gamma$ from a message set $\mathcal{M}$ to $\ca{X}^n$.
A decoder $\Pi$ 
is a POVM $\{\Pi_{m}\}_{m \in \mathcal{M}}$ on 
$\ca{K}_B^{\otimes n}$.
A pair of an encoder $\Gamma$ and a decoder $\Pi$ is called
a code $\Phi$. 
The decoding error probability is denoted by $\epsilon(W_B|\Phi)$.
The ratio $\frac{\log |\ca{M}|}{n}$ is called
the transmission rate and is denoted by $R(\Phi_n)$.

Here we measure the information leakage with a slightly different quantity from \eqref{CLP},
\begin{align}
\bar{d}(M;E)_{\tau_{ME}} &\coloneqq 
\norm{\tau_{ME}- P_\ca{M} \otimes\tau_{E}   }_{1}, \Label{eq:DefTrd}
\end{align}
where $\tau_{ME}$ is the joint state between Eve's system $E$ and
the message $M$ with the uniform distribution. 
Note that mutual information $I(M;E)$ as another frequently used criterion 
can be upper bounded by trace norm through Fannes' inequality.
When Alice's encode is $\Gamma$,
we denote the above value by $\bar{d}(M;E)[\Gamma] $.

A rate $R$ is said to be achievable for the wire-tap channel $(W_B,W_E)$
if there exists a sequence of codes $\Phi_n=(\Gamma_n,\Pi_n)$ such that 
when the time channel use $n$ goes to infinity,
the decoding error probability $\epsilon(W_B|\Phi_n)$
and information leakage 
$\bar{d}(M;E)[\Gamma_n] $
go to zero and the transmission rate 
$R(\Phi_n) $
goes to $R$.
The secrecy capacity for the wire-tap channel is defined as the supreme of all possible achievable rate, 
which is calculated as \cite{Devetak2005}
\begin{align}
	C(W_B,W_E) = \lim_{n \to \infty} \frac{1}{n} \max_{P_{TX^n}} \left[I(T;B^n) - I(T;E^n) \right].
\end{align}
As a corollary, the rate $I(X;B)-I(X;E)$ is always achievable.

To ensure that the transmitted message can be decoded reliably, an error correcting code for channel $W_B$ is needed. To keep the eavesdropper ignorant of the message, some randomization should be performed on the message. Hence, the modular coding scheme here is constructed as a concatenation of an inverse UHF family and an ordinary error correcting code. Such structure has been introduced in classical situations \cite{Hayashi2010}. 
In contrast to random coding \cite{Devetak2005,Hayashi2015} and ad hoc coding inspired by specific error correcting code \cite{Thangaraj2007,Mahdavifar2011,Klinc2011}, our scheme is more practical for implementation thanks to the modular structure. Also, our secrecy bound is improved compared with \cite{Hayashi2015}.

We take the following notations for the next parts. If a probability transition matrix $\Gamma:\ca{V}\rightarrow \ca{X}$ is applied to a random variable $V$, we use the symbol $\Gamma\circ P_V$ to denote the probability distribution of the transition output, and the symbol $\Gamma \times P_V$ to denote the joint distribution of both the input and the output. In other words, we define 
$\Gamma\circ P_V (x):= \sum_v P_V(v)\Gamma(x|v)$ and 
$\Gamma\times P_V (x,v):= P_V(v)\Gamma(x|v)$. 
Similarly, for the quantum channel $W_E$, we have the notations 
$ W_E \circ P_X := \sum_x P_X (x)W_{E}(x)$ and
$W_E \times P_X := \sum_x P_X(x) 
\proj{x}\otimes W_{E}(x)$.
Also, we define the cq channel $ W_E \circ \Gamma$ as
$ W_E \circ \Gamma (v) := \sum_x \Gamma (x|v)W_{E}(x)$. 
In addition, for any set ${\cal X}$, we denote the uniform distribution on the set ${\cal X}$ by $P_{{\cal X}}$.

\subsection{Modular code construction \label{MCC}}
The modular code is constructed based on an existing error correcting code. 
Before the construction, we introduce 
a UHF family $\{f_s\}_{s\in \ca{S}}$,
where $f_{S}$ is a function from a set $\ca{L}$ to another set $\ca{M}'$
and $S \in \ca{S}$ is a variable to identify the function
and is subject to the uniform distribution $P_{{\cal S}}$ on $\ca{S}$.

The function family $\{f_S\}_{S\in \ca{S}}$ is called 
a UHF family when the following condition holds;
\begin{description}
\item[(C1)]
The relation
\begin{align} \Label{eq:UHF}
\Pr \left[f_S(l) = f_S(l') \right] \le \frac{1}{\sM'}.
\end{align}
holds for any $ l \neq  l'\in \ca{L}$, where $\sM':=|{\cal M}'| $.
\end{description}
Specifically, we impose an additional balanced condition: 
\begin{description}
\item[(C2)]
The relation
\begin{align}
\abs{\{l\in \ca{L}:f_s(l)=m'\}} =\sL_2:= \frac{\sL_1}{\sM'} ,\Label{bala}
\end{align}
holds for any $s \in \ca{S}, m'\in\ca{M}'$, where $\sL_1 := \card{L}$. 
\end{description}
Throughout this paper we always mean the balanced version by UHF family.
The inverse of $f_s$ generates a random map.
That is, for a given $m' \in \ca{M}'$, 
we define the distribution $\Gamma[f_s]_{m'}$ as the uniform distribution 
on the set $\{l:f_s(l)=m\}$.

Suppose that our existing error correcting code is given as 
$(\ca{L},\{\Pi_l\}_{l\in \ca{L}})$, 
where $\ca{L}$ is a subset of channel input set 
$\ca{X}$
and $\{\Pi_l\}_{l\in \ca{L}}$ 
is a POVM for decoding $\ca{L}$.
We prepare 
a UHF family $\{f_s\}_{s\in \ca{S}}$ from $\ca{L}$ to $\ca{M}'$.
Before communication, Alice randomly chooses $S=s$, which will be shared via the public channel.
When Alice intends to send the message $m' \in \ca{M}'$,
she generates $L$ subject to the distribution $\Gamma[f_s]_{m'} $.
Therefore, when Alice intends to send $m' \in \ca{M}'$, Bob
receives the state
$W_B \circ \Gamma[f_s]_{m'}$.

The decoder at Bob's side is a  POVM $\Pi[f_s]$ with elements
\begin{align}
	\Pi[f_s]_m = \sum_{l: f_s(l)=m} \Pi_l \Label{eq:decoder}.
\end{align}
Then the encoder and the decoder constitute the code 
$\Phi[f_s] = (\Gamma[f_s],\Pi[f_s] )$.
The performance of a wire-tap code consists of error probability and secrecy. When the message $M$ is subjected to uniform distribution, we have the average error probability
\begin{align}
\epsilon(W_B|\Phi[f_s]) = 
 \sum_{m' \in \ca{M}'} \frac{1}{\sM'}\tr (1-\Pi[f_s]_{m'}) 
 W_B \circ \Gamma[f_s]_{m'}.
\end{align}
When we denote the decoding error probability of the code $\varphi$ by
$\epsilon(W_B|\varphi)$, we have
\begin{align}
\epsilon(W_B|\Phi[f_s]) \le \epsilon(W_B|\varphi).\Label{CKO}
\end{align}

For the information leakage, 
taking the average for $S$, we have
\begin{align}
\bar{d}(M';ES)[\Gamma[f_S]] &:= \norm{\tau_{M'ES}- P_{\ca{M}'} 
\otimes\tau_{ES}   }_1 \nonumber\\
&= \ex{S} \norm{\tau_{M'E|S} - P_{\ca{M}'}\otimes \tau_{E|S}}_1 , \nonumber\\
&= \ex{S} \bar{d}(M';E)[\Gamma[f_S]] , 
\Label{eq:DefTrd2}
\end{align}
where $\tau_{M'ES} = \sum_s P_S(s)\proj{s}\otimes W_E\circ 
\Gamma[f_s]\times P_{\ca{M}'}$ and $M'$ and $S$ are subjected to independent and uniform distribution. 
The quantity $\bar{d}(M';ES)$ is often useful for our analysis.


\subsection{Finite-length analysis for general wire-tap channel}
For our finite length analysis, we introduce 
\renyi mutual information as
\begin{align}
	I_{1+t}^{\uparrow}(A;B|\rho_{AB}) &\coloneqq D_{1+t}(\rho_{AB}\|\rho_A \otimes \rho_{B}),\\
	\tilde{I}_{1+t}^{\uparrow}(A;B|\rho_{AB}) &\coloneqq \tilde{D}_{1+t}(\rho_{AB}\|\rho_A \otimes \rho_{B}).
\end{align} 
Another version of \renyi mutual information is given as
\begin{align} 
	I_{1+t}^{\downarrow}(A;B|\rho_{AB}) &\coloneqq \inf_{\sigma_B\in \ca{D}(\ca{H}_B)} D_{1+t}(\rho_{AB}\|\rho_A \otimes \sigma_B), \\
	\tilde{I}_{1+t}^{\downarrow}(A;B|\rho_{AB}) &\coloneqq \inf_{\sigma_B\in \ca{D}(\ca{H}_B)} \tilde{D}_{1+t}(\rho_{AB}\|\rho_A \otimes \sigma_B) .
\end{align}

For the secrecy, we have the following theorem, which is shown in Appendix \ref{AppA1}.
\begin{theorem}
	\Label{thm:main}  
	Assume $W_E$ is a cq channel. 
	For any subset $\ca{L} \subset \ca{X}$, we choose a UHF family $\{f_s\}_{s \in \ca{S}}$ defined on $\ca{L}$. 
	Then, the wire-tap code $\Phi[f_s]$ with encoder $\Gamma[f_s]$ satisfies the following relation for the average information leakage. 
	\begin{align}
	\bar{d}(M;ES)[\Gamma[f_S]] &\le  \min_{0\le t \le 1} 2^{\frac{1-t}{1+t}}2^{\frac{t}{1+t}(-\log \sL_2 + \tilde{I}_{1+t}^{\downarrow}(X;E|W_E\times P_\ca{L}))} \Label{eq:Trd}.
	\end{align}
\end{theorem}

Since an error correcting code for a cq channel can always be transformed 
into a modular wire-tap code, 
we have the following proposition on error probability according to previous result \cite{Hayashi2007}.
\begin{proposition}\Label{PRO1}
	Assume $W_B$ is a cq channel. There exists an error correcting code 
	$\varphi=(\ca{L},\{\Pi_l\}_{l\in \ca{L}})$ such that
	\begin{align}
	\epsilon(W_B|\varphi) &\le 4 \min_{P_X,0\le t \le 1} 
	2^{t\left[ \log \sL_1 - I_{1-t}^{\uparrow}(X;B|W_B\times P_X) \right]} \Label{eq:error2}.
	\end{align}
\end{proposition}
When we choose a UHF family $\{f_s\}_{s \in \ca{S}}$ defined on a 
subset $\ca{L}$ chosen in the way as Proposition \ref{PRO1},
the combination of \eqref{CKO} and \eqref{eq:error2} implies the inequality
	\begin{align}
	\epsilon(W_B|\Phi[f_s]) &\le 4 \min_{P_X,0\le t \le 1} 
	2^{t\left[ \log \sL_1 - I_{1-t}^{\uparrow}(X;B|W_B\times P_X) \right]} \Label{eq:error}
	\end{align}
for any $s \in \ca{S}$.

Theorem 1 gives a single shot bound on information leakage. However, they are inconvenient to evaluate because of the dependence on the subset $\ca{L}$ in \eqref{eq:Trd}. The following corollary shows evaluable bound independent of $\ca{L}$ in the $n$-fold channel situation (see Appendix \ref{AppA2} for the proof). 

\begin{corollary} \Label{coro1}
When the cq channels $W_E$, $W_B$ take the $n$-fold form $W_E^{n}$, $W_B^n$, there exists a wire-tap code $\Phi[f_s]$ generated from $\varphi=(\ca{L},\{\Pi_l\}_{l\in \ca{L}})$ such that
\begin{align}
	\epsilon(W_B|\Phi[f_s] ) &\le 4 \min_{0\le t \le 1}
	 2^{t\left[ \log \sL_1 - n \max_{Q_X}
		I_{1-t}^{\uparrow}(X;B|W_B\times Q_X) \right]} \Label{eq:nferror}.
\end{align}
\end{corollary} 

\begin{corollary} \Label{coro1-2}
For any subset $\ca{L} \subset \ca{X}^n$, we choose a UHF family $\{f_s\}_{s \in \ca{S}}$ defined on $\ca{L}$.
	Then, the wire-tap code $\Phi[f_s]$ with encoder $\Gamma[f_s]$ satisfies the following relation for the average information leakage. 
\begin{align}
	\bar{d}(M';ES)[\Gamma[f_S]]
	&\le \min_{0\le t \le 1} 2^{\frac{1-t}{1+t}}2^{\frac{t}{1+t}
	(-\log \sL_2 + n\max_{Q_X}\tilde{I}_{1+t}^{\downarrow}(X;E|W_E\times Q_{X}))} \Label{eq:nfTD}
\end{align}
\end{corollary}

For any $0<t<1$ and arbitrarily small $\delta$, take $\log \sL_1 = n \max_{Q_X}
I_{1-t}^{\uparrow}(X;B|W_B\times Q_X) - n\delta$, 
$\log \sL = \max_{Q_X} \tilde{I}_{1+t}^{\downarrow}(X;E|W_E \times Q_X) + n\delta$, then any coding rate below 
$\max_{Q_X}I(X;B|W_E\times Q_X) - \max_{Q_X}I(X;E|W_E\times Q_X)$ is achievable by taking $t\to 0$.

\subsection{Finite-length analysis for symmetric wire-tap channel}\Label{SBA}
Next we discuss the symmetric channel case. A channel $W:\ca{X}\to \ca{D}(\ca{H})$ is symmetric
if the input set $\ca{X}$ is a group and 
there exist a unitary projective representation $\{U_x\}_{x \in \ca{X}}$ 
and 
a state $\rho_0 \in \ca{D}(\ca{H})$  
such that $W(x) = U_x \rho_0 U_x^\dagger$. 
When $W_E$ is symmetric, 
the security evaluation in our criterion automatically implies the semantic security\cite[Lemma 7]{Hayashi2015}.
Although Lemma 7 in \cite{Hayashi2015} showed the semantic security
by using the additive condition, i.e., the commutativity of $\ca{X}$,
this derivation uses only the above symmetric condition and  does not use 
the commutativity of $\ca{X}$.
In fact, while we will discuss 
a wire-tap channel $(W_B,W_E)$ for our analysis on private dense coding,
both channels $W_E$ and $W_B$ satisfy the symmetric condition.
As the following theorem, which is shown in Appendix \ref{A-thm:add}, 
we can simplify our upper bounds for the error probability and secrecy in the symmetric case. 
  
\begin{theorem}\Label{thm:add} 
When $W_E$ is a symmetric cq channel,
the upper bound in \eqref{eq:nfTD} is simplified to 
	\begin{align}
		\bar{d}(M';ES)[\Gamma[f_S]] \le \min_{0\le t \le 1} 2^{\frac{1-t}{1+t}}2^{\frac{t}{1+t}(-\log \sL_2 + n \tilde{I}_{1+t}^{\downarrow}(X;E|W_E\times P_{\ca{X}}))} \Label{eq:Trdadd}.
	\end{align}
When $W_B$ is a symmetric cq channel,
the upper bound in \eqref{eq:nferror} is simplified to 
\begin{align}
	\epsilon(W_B|\Phi[f_s] ) \le 4 \min_{0\le t \le 1}
	 2^{t\left[ \log \sL_1 - n 
	 	I_{1-t}^{\uparrow}(X;B|W_B\times P_{\ca{X}}) \right]} \Label{eq:nferror2}.
\end{align}
\end{theorem}

Theorem \ref{thm:add} shows that the rate 
\begin{align}
	R_* = I(X;B|W_E\times P_\ca{X}) - I(X;E|W_E \times P_\ca{X}) \Label{eq:RateQW}
\end{align} 
is achievable for our modular code in the symmetric case
while this achievability is known in existing studies \cite{Devetak2005,Cai2004,Hayashi2015}.
To see our advantage over existing studies, we focus on 
the exponential decreasing rate of information leakage for an encoder 
$\Gamma[f_S]$.
When the sacrificed rate $R_2 = \frac{\log \sL_2}{n}$ is fixed, 
$\bar{d}(M';ES)[\Gamma[f_S]]$ should decrease exponentially as $n\to \infty$. The exponential decreasing rate (exponent) is defined by
\begin{align}
e_d(R_s|W_E) \coloneqq \lim\limits_{n\to \infty} -\frac{1}{n} \log 
\bar{d}(M';ES)[\Gamma[f_S]].
\end{align}
Then \eqref{eq:Trdadd} yields 
\begin{align}
e_d(R_2|W_E)\ge  \max_{0\le t \le 1}\frac{t}{1+t}(R_2 - \tilde{I}_{1+t}^{\downarrow}(X;E|W_E\times P_{\ca{X}})).
\end{align}
The above lower bound of the exponent is strictly larger than the result 
\begin{align}
	e_d(R_2|W_E) \ge \max_{0\le t \le 1} \frac{t}{2} (R_2 - I_{1+t}^{\downarrow}(X;E|W_E\times P_X))
\end{align}
obtained in \cite{Hayashi2015} using random coding method.

Also, as shown in Appendix \ref{A-A7}, we have the following lemma.
\begin{lemma}\Label{A7}
When the cq wire-tap channel $(W_B,W_E)$ is degraded and both channels
$W_B,W_E$ are symmetric,
the secrecy capacity is $I(X;{\hat{B}})-I(X;{\hat{E}}) $, where $X$
express the random variable that takes values in the finite group $G$ according to the uniform distribution on $G$.
\end{lemma}

\section{Error verification} \Label{S6}
Now, we explain how the reliability can be realized by error verification \cite[Section VIII]{PhysRevA.81.012318}\cite{Hayashi2016-inspired}.
Assume that Alice sends the information $(M,Y)\in \ca{M}\times \ca{Y}$
via wire-tap code
and Bob obtains $(\hat{M},\hat{Y})\in \ca{M}\times \ca{Y}$.
$M$ and $Y$ are assumed to obey the uniform distribution on $\ca{M}$ and $ \ca{Y}$ independently.
They intend to check whether $M=\hat{M}$ holds
without leaking the information for $M$.
For this aim, Alice prepares 
a UHF family $\{g_{S'}\}:\ca{M} \times \ca{Y}\to \ca{Y}$, 
where $S'$ is the random seed to decide the hash function.
Here, we additionally impose the following condition;
\begin{description}
\item[(C3)]
For any $m \in \ca{M}$, $c \in \ca{Y}$, and $s' \in \ca{S}'$, 
there uniquely exists
$y(m,s',c) \in \ca{Y}$ such that
\begin{align}
c=g_{s'}(m,y(m,s',c)).
\end{align}
\end{description}
The condition (C3) implies the balanced condition (C2).
Then, Alice sends the random seed $S'$ and $C:=g_{S'}(M,Y)$ to Bob via the public channel.
When $g_{S'}(\hat{M},\hat{Y})=C $, Bob accepts his decoded message $\hat{M}$.
Otherwise, he aborts the protocol. 

Denote the length of $\ca{Y}$ by $\mathsf{t}$, i.e. $\mathsf{t} = \log \card{Y}$. 
In the following, 
we show that the relation 
\begin{align}
\sup_{m\ne \hat{m} }\Pr[\Ab^c|m,\hat{m}] 
\le 2^{-\mathsf{t}} \Label{CPA}
\end{align}
holds when $S'$ is independent of $M,Y,\hat{M},\hat{Y}$.
For this aim, it is sufficient to show the relation
\begin{align}
\Pr [g_{S'}(m,Y)=g_{S'}(\hat{m},\hat{Y})| M=m,\hat{M}=\hat{m}]\le 2^{-\mathsf{t}}
\end{align}
for any $m\ne \hat{m}\in \ca{M}$.
The above relation follows from the following relation;
The relation 
\begin{align}
\Pr [g_{S'}(m,y)=g_{S'}(\hat{m},\hat{y})]\le 2^{-\mathsf{t}}
\end{align}
for any $m\ne \hat{m}\in \ca{M}$ and $y,\hat{y} \in \ca{Y}$.
However, this relation follows from the definition of UHF.
Hence, we obtain the \eqref{CPA}, which is Reliability (S2).

The following lemma shows that the publicly shared variables for error verification 
give no information about the message $M$. 

\begin{lemma}\Label{LNM}
Assume that $M'=(M,Y) $ is subject to the uniform distribution
and $E'$ is a quantum system correlated to $M'$.
When $S'$ is an independent variable of other systems $M',E$,
and $C=g_{S'}(M,Y)$,
we have 
\begin{align}
d(M; E' S' C )\le d(M'; E' ).\Label{BMA}
\end{align}
\end{lemma}
This lemma is shown in Appendix \ref{ANL}.

\section{Detailed analysis for private dense coding}\Label{S7}
In this section, we provide the concrete protocol construction 
for general private dense coding setting
and derive the nonasymptotic and asymptotic performance of the protocol.
Then we show that the code can be practically implemented when certain conditions are satisfied.

\subsection{Protocol construction}\Label{S7-A}
We propose our concrete protocol for private dense coding $\text{PDC}(\tau_{ABE},\{U_g\}_{g\in G},\Lambda_A)$
by combining a wire-tap channel code and error verification.
Assume that $n$ copies of $\rho_{ABE}$ are given among Alice, Bob, and Eve.
Alice and Bob prepare an error correcting code 
	$\varphi=(\ca{L},\{\Pi_l\}_{l\in \ca{L}})$ for $n$ uses of cq-channel
	$g (\in G)\mapsto \Lambda_A( U_g \rho_{AB}U_g^\dagger )$.
For our efficient construction of our protocol, 
we choose a prime power $q$, which corresponds to the size of our finite field $\FF_q$
to be used. Then, we assume the condition for the code $\varphi$;
\begin{align}
|\ca{L}|=q^{n_1}. \Label{NAO}
\end{align}
If this condition does not hold,
we decrease the number of $ |\ca{L}|$ to satisfy this condition.
We choose a bijective map $\varphi_e$ from $\FF_q^{n_1}$ to $\ca{L}$.
We prepare the message set
$\ca{M}:=\FF_q^{n_2}$,
the set for covering variable $\ca{Y}:=\FF_q^{n_3}$,
and the message set for wire-tap code
$\ca{M}':=\ca{Y}\times \ca{M}=\FF_q^{n_2+n_3}$.

For $V\in \FF_q^{d_1+d_2-1}$,
we introduce the $d_1\times d_2$ Toeplitz matrix $T_{d_1,d_2}(V)$, 
which is defined as
\begin{align}
T_{d_1,d_2}(V)_{i,j}:=V_{i-j+d_2}.
\end{align}
We employ two UHF families $f_S:\FF_q^{n_1} \to \ca{M}'$
and $g_{S'}:\ca{M}'\to \ca{Y}$, where
$S \in \ca{S}:= \FF_q^{n_1-1}$
and $S' \in \ca{S}':= \FF_q^{n_2+n_3-1}$ are uniform random variables.
The UHF $f_S$ is defined as
\begin{align}
f_S(L)
=\left(
\begin{array}{c}
f_{S,1}(L) \\
f_{S,2}(L)
\end{array}
\right)
:=
(I, T_{n_2+n_3,n_1-(n_2+n_3) }(S))
\left(
\begin{array}{c}
L_1 \\
L_2
\end{array}
\right),\Label{E64}
\end{align}
where $L=(L_1,L_2)$, $L_1 \in \FF_q^{n_2+n_3}$, $L_2 \in \FF_q^{n_1-(n_2+n_3)}$
and 
$f_{S,1}(L)\in \ca{Y},f_{S,2}(L)\in \ca{M}$.
Similarly, the UHF $g_{S'}$ are defined with
$S' \in \ca{S}':= \FF_q^{n_2+ n_3-1}$ as
\begin{align}
g_{S'}(M,Y)
:=
(I, T_{n_3,n_2 }(S'))
\left(
\begin{array}{c}
Y \\
M
\end{array}
\right).\Label{E65}
\end{align}

When Alice intends to send the message $M\in\ca{M}$, 
Alice generates random variables
$S\in \ca{S}$,
$S'\in \ca{S'}$,
$Y\in \ca{Y}$, and 
$L_2\in \FF_q^{n_1-(n_2+n_3)}$
independently according to the uniform distribution.
Alice applies the encoder $\psi_S$ defined as \cite[Appendix A-B]{Vazquez-Castro}
\begin{align}
\psi_S(M,Y,L_2):=
\varphi_e
\left(
\left(
\begin{array}{cc}
I & -  T_{n_2+n_3,n_1-(n_2+n_3) }(S)\\
0 & I
\end{array}
\right)
\left(
\begin{array}{c}
M' \\
L_2
\end{array}
\right)
 \right),\Label{E66}
\end{align}
where $M'=(M, Y)^T$.
It is easy to verify that the encoder $\psi_s$ is an example of 
the encoder $\Gamma[f_s]_{m'}$ proposed in Section \ref{MCC}.
Then, based on an error correcting code $\varphi$ for cq-channel,
and integers $n_2$, $n_3$, prime power $q$,
we give our protocol $P(\varphi,n_2,n_3,q)$ as follows.

{\bf Potocol 4} 
\begin{algorithmic}
\STATE {\bf Encoding:}\quad 
To begin with, 
there are $n$ preshared states $\tau_{ABE}$
between Alice, Bob and Eve.
When Alice intends to send the message $M\in\ca{M}$, 
Alice generates random variables
$S\in \ca{S}$,
$S'\in \ca{S'}$,
$Y\in \ca{Y}$, and 
$L_2\in \FF_q^{n_1-(n_2+n_3)}$
independently according to the uniform distribution.
Alice chooses elements $(g_1g_2\ldots g_n):=\psi_S(Y,M,L_2)$.
Alice applies private dense coding operation 
$U_{g_i}$ on the $i$-th state and sends them to Bob.

\STATE {\bf Decoding 1:}\quad 
Bob applies measurement $\Pi=\{\Pi_{l}\}_{l \in \ca{L}}$
on the composite system $(\ca{H}_A\otimes \ca{H}_B)^{\otimes n}$
and obtains $\hat{L} \in \ca{L}$.
Bob acknowledges this decoding to Alice via the public channel.

\STATE {\bf Pubic communication from Alice to Bob:}\quad 
Alice sends the variables $S$, $S'$, and $C:=g_{S'}(M,Y)$ to Bob via
the public channel.

\STATE {\bf Decoding 2 \& Verification :}\quad 
When $g_{S'}\circ f_{S} \circ \varphi_e^{-1}(\hat{L})\neq C$, 
Bob aborts the protocol.
Otherwise, he recovers the message as
$\hat{M}:= f_{S,2} \circ \varphi_e^{-1}(\hat{L})$.
\end{algorithmic}

In the above protocol,
the part except for the error correcting code 
	$\varphi=(\ca{L},\{\Pi_l\}_{l\in \ca{L}})$
has calculation complexity $O(n\log n)$ due to the following reason.
Toeplitz matrix can be constructed as a part of circulant matrix. 
For example, the reference \cite[Appendix C-B]{T-H}
gives a method to give a circulant matrix.
Also, the reference \cite[Appendix C-A]{T-H} 
gives an algorithm for multiplication of a circulant matrix with 
calculation complexity $O(n\log n)$. 
Hence, if the error correcting part can be efficiently implemented,
this protocol can be efficiently implemented.

\begin{remark}
In the above protocol,
Alice needs to perform the public communication to Bob after
Bob receives the quantum states.
If Eve knows the variables $S'$ and $C=g_{S'}(M,Y)$
before Bob receives the quantum states,
Eve has a possibility that 
she can send the quantum state $\rho$ to Bob 
such that Bob's outcomes $\hat{M},\hat{Y}$ satisfy  
$C=g_{S'}(\hat{M},\hat{Y})$ and $\hat{M}=M$.
To avoid this risk, 
Alice needs to perform the public communication to Bob 
in this order.
\end{remark}

\subsection{Performance of our PDC protocol}
To apply the analysis of wire-tap channel in the evaluation of our PDC protocol, 
we have the following lemma, which is shown in Appendix \ref{A-lemma:AP}.
\begin{lemma} \Label{lemma:AP}
Given a PDC model $\text{PDC}(\tau_{ABE},\{U_g\}_{g\in G},\Lambda_A)$, 
we consider the two channels $W_{B}(x)= U_x \Lambda_A(\tau_{AB}) U_x^\dagger$ and 
$W_{E}(x) = U_x \tau_{AE} U_x^\dagger$.
Then, the information quantities can be expressed as
\begin{align}
	I_{1-t}^\uparrow (X;BB'|W_B\times P_{\ca{X}}) &= \log d_A - H_{1-t}^\downarrow(A|B|\Lambda_A(\tau_{AB})) \Label{eq:APDC1}\\
	\tilde{I}_{1+t}^{\downarrow} (X;AE|W_E\times P_{\ca{X}}) &= \log d_A - \tilde{H}_{1+t}^\uparrow(A|E|\tau_{AE}). \Label{eq:APDC2}
\end{align}
\end{lemma}

Using this lemma and Theorem \ref{thm:add},
as performance evaluation of our PDC protocol, we have the following theorem, which is shown in the end of this subsection.
\begin{theorem} \Label{thm:AP}
Given integers $n_2$, $n_3$, a prime power $q$,
and an error correcting code 
$\varphi=(\ca{L},\{\Pi_l\}_{l\in \ca{L}})$ such that $|\ca{L}|=q^{n_1} $ 
the protocol $P(\varphi,n_2,n_3,q)$ satisfies the following inequalities
\begin{align}
\epsilon_C(P(\varphi,n_2,n_3,q)) & \le \epsilon (\varphi) 
\Label{eq:errorA}\\
\epsilon_E(P(\varphi,n_2,n_3,q)) &\le 
\min_{0\le t \le 1} 
	2^{-\frac{1-t}{1+t}}2^{\frac{tn}{1+t}(- \frac{n_1-n_2-n_3}{n}\log q +\log d_A
	- \tilde{H}_{1+t}^\uparrow(A|E| \tau_{AE}))} \Label{eq:TrdA}\\
\epsilon_B(P(\varphi,n_2,n_3,q)) &\le q^{- n_3},\Label{eq:TrdY}
\end{align}
where $\epsilon (\varphi)$ is the decoding error probability of the error correcting code $\varphi$. 
\end{theorem} 

For 
a more concrete bound  on $\epsilon_C(P(\varphi,n_2,n_3,q))$,
we apply Proposition \ref{PRO1} to the case with 
$\log \sL_1= n_1 \log q$.
When $\varphi$ is an error correcting code given in Proposition \ref{PRO1}, we have
\begin{align}
\epsilon_C(P,n_2,n_3,q) &\le 4 \min_{0\le t \le 1} 
	2^{t n \left[ \frac{n_1}{n}\log q -\log d_A + H_{1-t}^\downarrow(A|B| \Lambda_A(\tau_{AB})) \right]} \Label{eq:errorA2}.
\end{align}
The RHS of \eqref{eq:errorA2}
follows from the application of \eqref{eq:APDC1} to 
the RHS of \eqref{eq:error2} in Proposition \ref{PRO1}.
Hence, \eqref{eq:errorA} implies \eqref{eq:errorA2}.

Therefore,
the key evaluations \eqref{eq:B1}, \eqref{eq:B2}, and \eqref{eq:B3}
in Subsection \ref{S2-A} are obtained as follows.
That is, \eqref{eq:errorA2}, \eqref{eq:TrdA}, and \eqref{eq:TrdY}
imply \eqref{eq:B1}, \eqref{eq:B2}, and \eqref{eq:B3}, respectively.
That is, the combination of Theorem \ref{thm:AP} and Proposition \ref{PRO1} guarantees the existence of a PDC protocol stated in Section \ref{S2-A}.

\begin{proofof}{Theorem \ref{thm:AP}}
When the error correcting code recovers $L$ correctly, 
the protocol does not abort.
Hence, combining \eqref{CKO} with the above fact, we obtain \eqref{eq:errorA}.
The relation \eqref{eq:TrdY} follows from \eqref{CPA}.

The cq-channel	 $g (\in G)\mapsto \Lambda_A( U_g \rho_{AE}U_g^\dagger )$
is symmetric,
Theorem \ref{thm:add} guarantees that
\begin{align}
d(M',EAS)\le \bar{d}(M',EAS)\le \min_{0\le t \le 1} 
	2^{-\frac{2t}{1+t}}2^{\frac{tn}{1+t}(- \frac{n_1-n_2-n_3}{n}\log q +\log d_A
	- \tilde{H}_{1+t}^\uparrow(A|E| \tau_{AE}))} \Label{eq:TrdAT},
	\end{align}
To derive the RHS of \eqref{eq:TrdAT}, we used \eqref{eq:APDC2}.
Applying Lemma \ref{LNM} to the case with $E'=EA$, 
we obtain \eqref{eq:TrdA} from \eqref{eq:TrdAT}.
Therefore, we obtain Theorem \ref{thm:AP}.
\end{proofof}

\subsection{Capacity formulas}
The private capacity characterizes the asymptotic performance
from a theoretical viewpoint.
The following theorem shows the equivalence between 
private capacity of private dense coding
and the secrecy capacity of corresponding wire-tap channel.
, which is proved in Appendix \ref{MMLA}.
\begin{theorem} \Label{lemma:APB}
For the PDC model  $\text{PDC}(\tau_{ABE},\{U_g\}_{g\in G},\Lambda_A)$ and
corresponding wire-tap channels $W_{B}(x)= U_x \Lambda_A(\tau_{AB}) U_x^\dagger$ and
$W_{E}(x) = U_x \tau_{AE} U_x^\dagger$, 
the private capacity of the PDC model
equals the secrecy capacity of $(W_B,W_E)$, i.e.,
\begin{align}
C(\tau_{ABE}, \{U_g\}_{g\in G},\Lambda_A)
=C(W_B,W_E).\Label{XL1}
\end{align}
\end{theorem}
Then, we have the following corollary.
\begin{corollary} \Label{A72}
When $\tau_{ABE}$ is a pure state, and $\tau_{AB}$ is maximally correlated,
we have
\begin{align}
C(\tau_{ABE}, \{U_g\}_{g\in G},\id_A)
=2 (H(\tau_{A,E})-H(\tau_E)) =2 H(A|E)_\tau=-2 H(A|B)_\tau
,\Label{XL2}
\end{align}
where $\id_A$ is the noiseless channel on $\ca{H}_A$.
\end{corollary}

This corollary can be shown as follows.
Appendix \ref{A1} shows that 
the wire-tap channel $(W_B,W_E)$ of the above case is degraded.
Due to Lemma \ref{A7} and the group symmetric condition,
the secrecy capacity of $(W_B,W_E)$ is the RHS of \eqref{XL2}.
Then, Theorem \ref{lemma:APB} guarantees \eqref{XL2}.

\subsection{Practical code construction with vector space over finite field}
Next, we show that 
the error correcting part $\varphi$ can be efficiently implemented,
under the following conditions (B1) and (B2) introduced in Section \ref{S2}
because the above subsection shows that other parts can be efficiently implemented.
\begin{description}
	\item[(B1)]
	The group $G$ forms a vector space $\ca{X}$ over a finite field $\FF_q$.
	\item[(B2)] The states $\{ U_x \Lambda_A( \tau_{AB}) U_x^\dagger\}_{x \in \ca{X}}$
	are commutative with each other.
\end{description}

To discuss the calculation complexity for the error correcting code,
it is sufficient to consider the case when 
the channel from Alice to Bob is $\Lambda_A$, i.e., is not intercepted by Eve.
Due to (B2), we can choose a basis $\{|e_\omega\rangle\}_{\omega\in\Omega}$
on $\mathcal{H}_{B'}\otimes \mathcal{H}_B$ that commonly diagonalizes
$U_x \Lambda_A(\tau_{AB}) U_x^\dagger$ for all $x \in \ca{X}$.
We denote the measurement corresponding to this basis
by $\{\Pi_\omega\}_{\omega\in\Omega}$, 
and define 
the distribution 
$P_{\Omega}(\omega):= \tr \Pi_\omega\Lambda_A(\tau_{AB})$.
We obtain the classical channel
$W^c(\omega|x):= \tr \Pi_\omega U_x \Lambda_A(\tau_{AB}) U_x^\dagger$.
Without any information loss, Bob's decoding can be reduced to 
the application of classical decoding for the classical channel $W^c$
to the outcomes via the measurement $\{\Pi_\omega\}_{\omega\in\Omega}$.

Since
the density matrix $U_x \Lambda_A(\tau_{AB}) U_x^\dagger$
has the same eigenvalue as $ \Lambda_A(\tau_{AB}) $ including the multiplicity,
there exists a permutation $\pi_x$ on $\Omega$ such that
$P_{\Omega}(\pi_x (\omega))= \tr \Pi_\omega U_x 
\Lambda_A(\tau_{AB}) U_x^\dagger$.
Since $\pi_x \pi_{x'}= \pi_{x + x'}$, the relation
$W^c(\omega|x)=P_{\Omega}(\pi_x (\omega))$
implies that the channel $W^c$ is a symmetric channel.

In this symmetric setting, we can choose an error correcting code 
as a linear subspace $\mathcal{L} \subset \ca{X}^n$ on $\FF_q$.
Such an error correcting code $\mathcal{L}$ can be constructed by using 
LDPC codes \cite{Thangaraj2007} or polar codes \cite{Mahdavifar2011}. 
It has been demonstrated that polar codes can achieve channel capacity 
for any discrete symmetric channels with 
a sufficiently small decoding error probability $\epsilon_C$ such that
the encoder $\phi_e$ and the decoder $\phi_d$ have 
calculation complexity $O(n\log n)$ \cite{Arikan2009,Sasoglu2009,Mori2010,Mori2014}.

\section{Conclusion and discussion}\Label{S8}
We have formulated the framework of private dense coding to realize quantum secure direct communication.
While this method requires preshared quantum state,
it works even when noise exists.
This method guarantees the secrecy against the eavesdropper, Eve, 
even when Eve intercepts the quantum system transmitted by Alice.
To cover the finite-length effect, we have derived a formula for 
the amount of information leakage dependently of the block length of our code and the sacrificed rate.
When the channel to Eve is symmetric,
this formula has better bound (Theorem \ref{thm:add})
than an existing bound \cite[(78)]{Hayashi2015}.

In our method, Alice's encoding operation is limited to unitary operation given as 
a projective unitary representation of a group $G$.
Also, when a certain commutative condition holds,
we have shown that Bob' decoding measurement can be restricted to 
a special measurement over a single joint system between Bob's receiving system and 
Bob's local system.
That is, Bob does not need to make any measurement across multiple joint systems.
Further, when the group is a vector space over a finite field,
we have proposed a practical coding method, whose encoding and decoding operations
have the calculation complexity $O(n \log n )$, where $n$ is the block length.

We have applied our results to the case when Alice's encoding operation is limited to the
Weyl-Heisenberg representation.
Although this case is a similar to the case studied by preceding papers \cite{Wang2005},
we have derived a security formula for the information leakage 
in the finite-length setting.
In this setting, to cover an unknown preshared state,
we have proposed another new protocol that contains estimation process of the preshared state.

There are many protocols and experiments 
that fit into our setting \cite{Deng2003,Wang2005,Zhang2017,Zhu2017,Qi2021,Wang2018}.
However, the rigorous performance in a practical situation is unclear.
Our results reduce the gap between the theoretical performance 
and experimental setting.
For a realistic application, we need to combine the error estimation and our 
evaluation for information leakage \eqref{ANO}. Such evaluation is a future study.
This paper assumes that the preshared state is $n$-fold i.i.d condition.
However, the realistic case does not necessarily satisfy this condition.
Removing this condition is another future topic.

Unfortunately, this study have not analytically derived the asymptotically tight transmission rate, which is called the capacity.
As stated in Corollary \ref{A72},
our analytically obtained transmission rate is tight when $\tau_{ABE}$ is a pure state, and $\tau_{AB}$ is maximally correlated.
This problem was inherited by the following property of wire-tap channel.
When the wire-tap channel model is degraded,
$\max_{P_X} I(X;B)-I(X;E)$ is the capacity, i.e., the optimal transmission rate. 
Otherwise, this value is not the capacity in general and the analytical derivation of 
the capacity is an open problem.
Hence, to resolve this problem, it is needed to derive 
the capacity for a wire-tap channel model under the symmetric condition.
It is another future problem.
In addition, recently, the one-step QSDC protocol was proposed \cite{Zhou2022,Sheng2022}.
Since this protocol has a form different from our private dense coding,
we cannot directly apply our result to this problem.
Hence, the analysis on this problem is another future study.

\section*{Acknowledgment}
The works reported here
were supported in part by Guangdong Provincial Key Laboratory (Grant No. 2019B121203002),
Tsinghua University Initiative Scientific Research Program,
and the National Natural Science Foundation of China (Grant No. 62171212).

\appendix

\section{Maximally correlated state}\Label{A1}
In this appendix, we show that
the cq wire-tap channel $W_B,W_E$ is degraded
when $\Lambda_A$ is the noiseless channel, $\tau_{ABE}$ is a pure state, and $\tau_{AB}$ is maximally correlated.

We choose bases $|v_j^A\rangle$ and $|v_j^B\rangle$ on
$\mathcal{H}_A$ and $\mathcal{H}_B$ such that
$\tau_{AB}=\sum_{j,j'}a_{j,j'}|v_j^A,v_j^B\rangle
\langle v_{j'}^A,v_{j'}^B|$.
We diagonalize
$\tau_{AB}$ as
$\tau_{AB}=
\sum_{j} s_j |u_j\rangle \langle u_j|
$ and
$|u_j\rangle= \sum_{j'} b_{j,j'}|v_{j'}^A,v_{j'}^B\rangle$,
where $( b_{j,j')}$ is a unitary matrix.
Then, by choosing a basis $|v_j^E\rangle $,
the pure state $\tau_{ABE}$ is written as
$\sum_{j}\sqrt{s_j} |u_j\rangle|v_j^E\rangle$.
We choose $t_{j'}$ and normalized vectors $(c_{j|j'})_j$ for $j'$
as $t_{j'}:= \sum_{j} s_j |b_{j,j'}|^2$ and $
\sqrt{s_j}b_{j,j'}= \sqrt{t_{j'}} c_{j'|j}$.
We define $|u_{j'}^E\rangle:= \sum_{j}c_{j|j'}|v_{j}^E\rangle$.
Hence, we have
$\sum_{j}\sqrt{s_j} |u_j\rangle|v_j^E\rangle
=\sum_{j,j'}\sqrt{s_j}b_{j,j'}|v_{j'}^A,v_{j'}^B\rangle|v_j^E\rangle
=\sum_{j'}\sqrt{t_{j'}}|v_{j'}^A,v_{j'}^B\rangle|u_{j'}^E\rangle$, which implies that
\begin{align}
	\tau_{AE}=
	\sum_{j'} 
	t_{j'}
	|v_{j'}^A ,u_{j'}^E\rangle\langle v_{j'}^A ,u_{j'}^E|.
\end{align}
We define the TP-CP map $\Gamma$ as
\begin{align}
	\Gamma( \rho):= \sum_{j'} 
	\langle v_{j'}^B|\rho |v_{j'}^B\rangle  
	|u_{j'}^E\rangle\langle u_{j'}^E|.
\end{align}
Then, we have
\begin{align}
	& \Gamma( \tau_{AB})= 
	\sum_{j'} 
	\langle v_{j'}^B|\tau_{AB} |v_{j'}^B\rangle  
	|u_{j'}^E\rangle\langle u_{j'}^E|\nonumber\\
	=&
	\sum_{j'} 
	t_{j'}
	|v_{j'}^A ,u_{j'}^E\rangle\langle v_{j'}^A ,u_{j'}^E|
	=\tau_{AE}.
\end{align}
Since $U_g$ acts only on $\mathcal{H}_A$,
the cq wire-tap channel $W_B,W_E$ is degraded.


\section{Bob receives Bell diagonal state}\Label{app:Bell}
Here, we show that Bob's received state is Bell diagonal in the setting $\text{PDC}(\ket{\Psi},\{\mbf{W}(x,z)\}, \Lambda[\tilde{P}_{XZ}]_{A})$. When Alice's operation is $\mbf{W}(x,z)$,
Bob's state on ${\cal H}_{B}\otimes {\cal H}_{B'}$
is
\begin{align}
 &\Lambda[\tilde{P}_{XZ}]_A 
(\mbf{W}(x,z)
 \Lambda[{P}_{XZ}]_A(|\Phi\rangle \langle \Phi|) \mbf{W}(x,z)^\dagger)\nonumber \\
=&\mbf{W}(x,z)
( \Lambda[\tilde{P}_{XZ}]_A \circ
 \Lambda[{P}_{XZ}]_A(|\Phi\rangle\langle \Phi|) )\mbf{W}(x,z)^\dagger \nonumber\\
=&\mbf{W}(x,z)
 \Lambda[\tilde{P}_{XZ}*{P}_{XZ}]_A(|\Phi\rangle\langle \Phi|) \mbf{W}(x,z)^\dagger \nonumber\\
=&
 \Lambda[F_{x,z}[ \tilde{P}_{XZ}*{P}_{XZ}] ]_A
 (|\Phi\rangle\langle \Phi|) \nonumber\\
 =& \rho[F_{x,z}[ \tilde{P}_{XZ}*{P}_{XZ}]],
 \end{align}
where the distributions 
$\tilde{P}_{XZ}*{P}_{XZ}$ and $F_{x,z}[P_{XZ}]$ on 
$\mathbb{F}_p^{2}$
are defined as
\begin{align}
\tilde{P}_{XZ}*{P}_{XZ}(x,z) &:= \sum_{x'z'}
\tilde{P}_{XZ}(x',z') {P}_{XZ}(x-x',z-z'), \\
F_{x,z}[P_{XZ}](x',z')&:=P_{XZ}(x'-x,z'-z),
\end{align}
and the density matrix 
$ \rho[P_{XZ}]$ is defined as
\begin{align}
 \rho[P_{XZ}]:= \sum_{x,z}P_{XZ}(x,z)
\mbf{W}(x,z) |\Phi\rangle \langle \Phi|\mbf{W}(x,z)^\dagger.
\end{align}

\section{Proof of Lemma \ref{lemma:PI1}}\Label{P-lemma:PI1}
For the setting  
$\text{PDC}(\omega_{ABE},\{\mbf{W}(x,z)\}, \id_{A})$, 
recall that 
$\omega_{ABE} = \Lambda[\tilde{P}_{-X,Z}]_B (\ket{\Psi}\bra{\Psi})$, where 
\begin{align}
	\ket{\Psi}_{ABE} = \frac{1}{\sqrt{d}}\sum_{x,z}
	\sqrt{P(x,z)} \mbf{W}_A(x,z)\ket{\Phi}_{AB} \ket{x,z}_E 
\end{align}
is the purification of $\rho[P_{XZ}]$.
Then the reduced density matrices are 
\begin{align}
	\omega_{AE} &=  \sum_{j}\frac{1}{d} \Proj (\sum_{x,z}\sqrt{P(x,z)}\mbf{W}(x,z)\ket{j}_A\ket{x,z}_E), \nonumber\\
   \omega_E &= \sum_{x,z} P_{XZ}(x,z)\ket{x,z}_E \bra{x,z},\nonumber \\
   \omega_{AB} &=  \rho[ \tilde{P}_{XZ}*{P}_{XZ}], \nonumber\\
   \omega_B &= \rho_{B,\mix}.
\end{align}

The quantities for asymptotic rate are as follows:
\begin{align*}
	H(A|E)_\omega &= H(\omega_{AE}) - H(\omega_E) \\
    &= \log d_A - H(XZ|P_{XZ}), \\
    H(A|B)_\omega &=  H(\omega_{AB}) - H(\omega_B) \\
    &=  H(XZ|\tilde{P}_{XZ}*P_{XZ}) - \log d_A.
\end{align*}

The quantities for finite analysis are as follows:
\begin{align*}
	\tilde{H}_{1+t}^\uparrow(A|E)_\omega
	&\ge \tilde{H}_{1+t}^\downarrow(A|E)_\omega \\
	&= - \tilde{D}_{1+t}(\omega_{AE}\|\omega_E) \\
	&= -\frac{1}{t} \log \tr \sandwc{\omega_{AE}}{\omega_E}  \\
	&=  - \frac{1}{t} \log \frac{1}{d_A^s} \left(\sum_{x,z}P_{XZ}(x,z)^{\frac{1}{1+t}}\right)^{1+t} \\
	&= \log d_A - H_{\frac{1}{1+t}}^\downarrow(P_{XZ}), \\
	H_{1-t}^\downarrow(A|B)_\omega &= - D_{1-t}(\omega_{AB}\|\omega_B) \\
	&= \frac{1}{t} \log \tr \rho[ \tilde{P}_{XZ}*{P}_{XZ}] ^{1-t} \rho_{B,\mix}^{s} \\
	&= - \log d_A + H_{1-t}^\downarrow (\tilde{P}_{XZ}*{P}_{XZ}). 
\end{align*}

\section{Estimation of Bell diagonal state}\Label{MKA}
\subsection{Case with Bell diagonal state}
We consider the state estimation on the composite system 
${\cal H}_A\otimes {\cal H}_B$
by using local measurements
when the unknown state is given as 
a Bell diagonal state $\rho[P_{XZ}]$, which is defined in 
\eqref{BOR}.
Here, we consider only the case when $d$ is a prime $p$ 
and the arithmetics is performed in $\FF_p$ in the following part.
First, we prepare the relation;
\begin{align}
\mbf{W}(x,z)\mbf{W}(x',z')=\omega^{x'z-xz'}
\mbf{W}(x',z')\mbf{W}(x,z).\Label{BK3}
\end{align}

For $k,l $, we measure the marginal distribution
$P_{lX-kZ}$ as follows.
We say that the following measurement is $M(lX-kZ)$. 
Alice measures $\mbf{W}(k,l) = \mbf{X}^k \mbf{Z}^l$
and obtains the outcome $\omega^{Y}$.
Bob measures $(\mbf{X}^k \mbf{Z}^l)^T=\mbf{Z}^{l} \mbf{X}^{-k} $,
and obtains the outcome $\omega^{\bar{Y}}$.
We denote the eigenvector of $\mbf{X}^k \mbf{Z}^l$ with eigenvalue $\omega^j$ on the system $\mathcal{H}_A$
by $|j\rangle_{\mbf{X}^k \mbf{Z}^l,A}$.
We define $|j\rangle_{ \mbf{Z}^{l}\mbf{X}^{-k},B}$ in the same way.
We have
\begin{align}
|\Phi\rangle= \sum_{j=0}^{p-1}\frac{1}{\sqrt{d}} 
|j\rangle_{\mbf{X}^k \mbf{Z}^l,A}
|j\rangle_{ \mbf{Z}^{l}\mbf{X}^{-k},B}\Label{BK1},
\end{align}
and \eqref{BK3} implies
\begin{align}
\mbf{W}(k,l) \mbf{W}(x,z) |j\rangle_{\mbf{X}^k \mbf{Z}^l,A}
=\omega^{xl-zk}\mbf{W}(x,z) \mbf{W}(l,k)  |j\rangle_{\mbf{X}^k \mbf{Z}^l,A}
=\omega^{xl-zk+j}\mbf{W}(x,z)   |j\rangle_{\mbf{X}^k \mbf{Z}^l,A}.\Label{BK2}
\end{align}
That is, $\mbf{W}(x,z) |j\rangle_{\mbf{X}^k \mbf{Z}^l,A}$
is a constant times of $|xl-zk+j\rangle_{\mbf{X}^k \mbf{Z}^l,A}$.
Hence, when we measure operator $\mbf{X}^k\mbf{Z}^l$ for state $\mbf{W}(x,z) \ket{\Phi}$,
the outcome $\omega^{Y-\bar{Y}}$ is $\omega^{ xl-zk}$.

We consider that the state $\rho[P_{XZ}]$ is generated by 
applying the operator $\mbf{W}(x,z)$ to the state 
$|\Phi\rangle$ with probability $P_{X,Z}(x,z)$.
When $\mbf{W}(x,z)$ is applied,
the outcome $\omega^{Y-\bar{Y}}$ is $\omega^{ xl-zk}$.
That is, $Y-\bar{Y}$ obeys the distribution
$P_{lX-kZ}$.
We define $E[ \omega^{l X-k Z } ]:=\sum_{s=0}^{p-1} \omega^s P_{l X-k Z}(s)$.
We define the equivalent relation $ (x,z) \sim (x',z')$ in $\mathbb{F}_p^2$
as  $(x,z) = (a x',az') $ with an element $a \in \mathbb{F}_p$.
Since $P_{lX-kZ}(as)=P_{a(lX-kZ)}(s) $ and 
$(\mathbb{F}_p^2\setminus \{0\})/ \sim =\{ [(1,0)], [(1,1)],\ldots, [(1,p-1)], [(0,1)]  \}$,
we can calculate $E[ \omega^{l X-k Z } ]$ 
only by measuring 
$P_{X}, P_{X+Z}, \ldots, P_{X+(p-1)Z}$, and $P_Z$, which requires $p+1$ types of measurements.
\begin{lemma}\Label{LL1}
We have the following relation;
\begin{align}
P_{X,Z}(l,j)= \frac{1}{d^2}\sum_{b,z}\omega^{-jl+bl-jz}E[ \omega^{(b-j) X+z Z } ].
\end{align}
\end{lemma}
For $z\neq 0$, 
$E[ \omega^{(b-j) X+z Z } ] $ can be calculated from the distribution
$P_{   \frac{(b-j)}{z}X+Z}$, and
$E[ \omega^{(b-j) X } ] $ can be calculated from the distribution $P_{X}$.
Hence, 
from $p+1$ distribution
$P_{X}, P_{X+Z}, \ldots, P_{X+(p-1)Z}$, and $P_Z$, 
we derive the distribution $P_{X,Z}$.

\begin{proof}
To show Lemma \ref{LL1}, 
we consider the linear space ${\cal V}$ of complex functions on $\mathbb{F}_p^2$ and its dual space ${\cal V}^*$.
We define $e_{(x,z)} \in {\cal V}^*$
as $e_{(x,z)}(f):= f(x,z) \in \mathbb{C} $
for $f \in {\cal V}$.
Then, we define the following function 
${\cal F}$
from ${\cal V}^*$
to $p \times p$ complex matrices;
\begin{align}
{\cal F}[ e_{(x,z)} ]:= 
(|x\rangle_{\mbf{X}} ~_\mbf{X}\langle x|)
(|z\rangle_{\mbf{Z}} ~_\mbf{Z}\langle z|),
\end{align} 
We extend $E[ \omega^{k X+l Z } ]$ as an element 
of ${\cal V}^*$ as
\begin{align}
E[ \omega^{k X+l Z } ](f):=
\sum_{(x,z)} \omega^{kx+lz}f(x,z).
\end{align} 

Hence, it is sufficient to show that ${\cal F} $ is invertible and
the following relation holds,
\begin{align}
(|l\rangle_{\mbf{X}} ~_\mbf{X}\langle l|)
(|j\rangle_{\mbf{Z}} ~_\mbf{Z}\langle j|)
=
\frac{1}{p^2}\sum_{b,z}\omega^{-jl+bl-jz}
{\cal F}[E[ \omega^{(b-j) X+z Z } ]].\Label{BLA}
\end{align}

We have
\begin{align}
&{\cal F}[E[ \omega^{l X-k Z } ]]=
\sum_{j=0}^{d-1} \sum_{(x,z): lx-kz=j}\omega^{j}
(|x\rangle_{\mbf{X}} ~_\mbf{X}\langle x|)
(|z\rangle_{\mbf{Z}} ~_\mbf{Z}\langle z|) \nonumber\\
=&
\sum_{x,z}\omega^{lx-kz}
(|x\rangle_{\mbf{X}} ~_\mbf{X}\langle x|)
(|z\rangle_{\mbf{Z}} ~_\mbf{Z}\langle z|) \nonumber\\
=&
(\sum_x \omega^{lx} |x\rangle_{\mbf{X}} ~_\mbf{X}\langle x|)
(\sum_s \omega^{-kz }|z\rangle_{\mbf{Z}} ~_\mbf{Z}\langle z|) 
= \mbf{X}^{l}\mbf{Z}^{-k} .
 \end{align} 
Since the set $\{\mbf{X}^{l}\mbf{Z}^{-k}\}_{(l,k)} $ spans 
the set $d \times d$ matrices,
the map ${\cal F} $ is invertible.

In fact, we have
\begin{align}
&(|l\rangle_{\mbf{X}} ~_\mbf{X}\langle l|)
(|j\rangle_{\mbf{Z}} ~_\mbf{Z}\langle j|) \nonumber\\
=&
\frac{\omega^{-j l}}{\sqrt{d}}
|l\rangle_{\mbf{X}} ~_\mbf{Z}\langle j| \nonumber\\
=&
\frac{\omega^{-j l}}{\sqrt{p}}
\sum_{b=0}^{p-1} \frac{\omega^{b l}}{\sqrt{p}}
|b\rangle_{\mbf{Z}} ~_\mbf{Z}\langle j| \nonumber\\
=&
\frac{\omega^{-j l}}{p}
\sum_{b=0}^{p-1} \omega^{b l}\frac{1}{p}
\sum_{z=0}^{p-1} \omega^{-jz}
\mbf{X}^{b-j}\mbf{Z}^{z}\nonumber\\
=&\frac{1}{p^2}\sum_{b,z}\omega^{-jl+bl-jz}
{\cal F}[E[ \omega^{(b-j) X+z Z } ]],
\end{align}
which implies \eqref{BLA}.
\end{proof}

\subsection{Case with general state}
Even when $\rho_{AB}$ is not a generalized Bell diagonal state,
the resultant state $T(\tau_{AB})$ of discrete twirling \eqref{BAP}
is a generalized Bell diagonal state.

When the state is given as the twirled state $T(\tau_{AB})$ and 
the measurement  $M(lX-kZ)$ is applied, the probability with the outcome $y$ is
\begin{align*}
&\sum_{j=0}^{p-1}
~_{\mbf{X}^k \mbf{Z}^l,A}\langle j+y|
~_{ \mbf{Z}^{-l}\mbf{X}^k,B}\langle j|
\frac{1}{d^2} \\
&\quad \Big(\sum_{x,z} 
(\mbf{W}(x,z)_A \otimes \mbf{W}(x,z)_B^T) \tau_{AB}
(\mbf{W}(x,z)_A \otimes \mbf{W}(x,z)_B^T)^\dagger\Big)
|j+y\rangle_{\mbf{X}^k \mbf{Z}^l,A}
|j\rangle_{ \mbf{Z}^{-l}\mbf{X}^k,B}
\\
=
&\sum_{j=0}^{p-1}
\frac{1}{d^2}\sum_{x,z} 
~_{\mbf{X}^k \mbf{Z}^l,A}\langle -xl+zk+j+y|
~_{ \mbf{Z}^{-l}\mbf{X}^k,B}\langle -xl+zk+j|
\tau_{AB} \\
&\quad |-xl+zk+j+y\rangle_{\mbf{X}^k \mbf{Z}^l,A}
 |-xl+zk+j\rangle_{ \mbf{Z}^{-l}\mbf{X}^k,B}\\
=
&\sum_{j=0}^{p-1}
~_{\mbf{X}^k \mbf{Z}^l,A}\langle j+y|
~_{ \mbf{Z}^{-l}\mbf{X}^k,B}\langle j|
\tau_{AB}
|j+y\rangle_{\mbf{X}^k \mbf{Z}^l,A}
|j\rangle_{ \mbf{Z}^{-l}\mbf{X}^k,B}.
\end{align*}
That is, the distribution of the outcome with the input state 
\eqref{BAP} is the same as 
the distribution of the outcome with the input state $\tau_{AB}$.
Hence, when $ \eqref{BAP}$ is given as $\rho[P_{XZ}]$,
the distribution $P_{lX-kZ}$ can be estimated by applying 
the measurement $M(lX-kZ)$ to the state $\tau_{AB}$.

Note that the twirling operation needs public communication about the  choice $(x,z)$.
If they apply the twirling operation, Eve can perfectly recover 
the environment of the twirled state $T(\tau_{AB})$.
Therefore, we can analyze the twirled state $T(\tau_{AB})$ for secrecy without applying it.
To conclude, we show that we do not need to apply the twirling operation.

\section{Proofs in Section \ref{S5} \Label{AppA}}

This appendix gives the proof details in Section \ref{S5}.

\subsection{Proof of Theorem \ref{thm:main}}\Label{AppA1}
For a cq stare $\rho_{L,E}= 
\sum_{l \in \ca{L}} |l\rangle \langle l| \otimes X_{l}$ and
a function $f:\ca{L}\to \ca{M}'$, we define
the cq state $f(\rho_{L,E}):=
\sum_{l \in \ca{L}} |f(l)\rangle \langle f(l)| \otimes X_{l}$.
Recall the expression of $\bar{d}(M';E S)$ in \eqref{eq:DefTrd}, where
\begin{align}
	\tau_{M'E|S=s} 
	&= \sum_{m'} \frac{1}{\sM'} \ket{m'}\bra{m'}\otimes \sum_{l\in f_s^{-1}(m')}\frac{1}{\sL_2}W_{E}(l) \notag \\
	&= f_s(W_E\times P_\ca{L}), \notag\\
	\tau_{E|S=s} &= \tr_{M'} \tau_{M'E|S=s} = W_E \circ P_\ca{L}. \notag
\end{align}
Then, we introduce the privacy amplification lemma \cite{Dupuis2021}.
\begin{lemma}
	$\tau_{LE} \in \ca{D}(\ca{H}_L\otimes \ca{H}_E)$ is classical on $L$, $\{f_S\}:\ca{L}\rightarrow \ca{M}'$ is a UHF family, then
	\begin{equation}
		\ex{S} \norm{f_S (\tau_{LE}) - P_{\ca{M}'}\otimes \tau_E}_1 \le 2^{\frac{1-t}{1+t}} 2^{\frac{t}{1+t}(\log \card{M'} -  \tilde{H}_{1+t}^\uparrow(L|E)_\tau)}.
	\end{equation}
\end{lemma}
\eqref{eq:Trd} follows immediately by using the above lemma.
\begin{equation}
	\begin{aligned}
		d(M';ES) &= \ex{S} \norm{\id_E\otimes f_S(W_E\times P_\ca{L})-P_{\ca{M}'} \otimes W_E\circ P_\ca{L}}_{1} \\
		&\le 2^{\frac{1-t}{1+t}} 2^{\frac{t}{1+t}(\log \sM' - \sup_{\sigma_E} \tilde{H}_{1+t}^\uparrow(L|E|W_E\times P_{\ca{L}}))} \\
		&= 2^{\frac{1-t}{1+t}}2^{\frac{t}{1+t}(-\log b + \tilde{I}_{1+t}^{\downarrow}(L;E|W_E\times P_ {\ca{L}}))}.
	\end{aligned}
\end{equation}

\subsection{Proof of Corollaries \ref{coro1} and \ref{coro1-2}}\Label{AppA2}
The following lemma would be useful in the proof.
\begin{lemma} \Label{lemma:nf}
	For $n$-fold channel $W_B^n$, the following inequalities hold for 
	$t>-1$ and $t>-\frac{1}{2}$ respectively.
	\begin{align} 
		\max_{Q_{X^n}} I_{1+t}^{\downarrow}(X^n;B^n|W_B^n\times Q_{X^n}) &= n \max_{Q_X} I_{1+t}^{\downarrow}(X;E|W_B \times Q_X) \Label{eq:ProI} \\
		\max_{Q_{X^n}} \tilde{I}_{1+t}^{\downarrow}(X^n;B^n|W_B^n\times Q_{X^n}) &= n \max_{Q_X} \tilde{I}_{1+t}^{\downarrow}(X;B|W_B \times Q_X) \Label{eq:ProIG}.
	\end{align}
\end{lemma}
\begin{proofof}{Lemma \ref{lemma:nf}}
The equation \eqref{eq:ProI} for Petz version has been shown in \cite{Ogawa1999}, 
so we focus on \eqref{eq:ProIG}. 
Before the proof of Lemma \ref{lemma:nf}, we prepare the following proposition.
\begin{proposition} \Label{lemma:P42}
	(Proposition 4.2 of \cite{Mosonyi2017}) For $t \ge -\frac{1}{2}$, we have
	\begin{equation}
		\sup_{Q_X} \tilde{I}_{1+t}^{\downarrow}(X;E|W\times Q_X) = \inf_{\sigma} \sup_{x\in \ca{X}} \tilde{D}_{1+t}(W(x)\|\sigma).
	\end{equation}
\end{proposition}

Then, we define the quasi-entropy for convenience,
\begin{align}
	\Xi_{1+t}(\rho\|\sigma) \coloneqq \tr  \sandwc{\rho}{\sigma}. \Label{eq:DefQuasi}	
\end{align}
For $t>-\frac{1}{2}$ and any probability distribution on $\ca{X}^n$, we have

\begin{align}
	& \tilde{I}_{1+t}^{\downarrow}(X^n;E^n|W^{\otimes n}\times Q_{X^n}) \notag \\
	& = \inf_{\omega \in \ca{D}(\ca{H}_E^{\otimes n})} \frac{1}{t}\log \sum_{x^n\in \ca{X}^n}Q_{X^n}(x^n)\Xi_{1+t}(W^{\otimes n}(x^n)\|\omega)  \Label{eq:appnf1}\\
	&\le \inf_{\sigma\in \ca{D}(\ca{H}_E)} \frac{1}{t}\log \sum_{x^n\in \ca{X}^n}Q_{X^n}(x^n)\Xi_{1+t}(W^{\otimes n}(x^n)\|\sigma^{\otimes n})   \notag \\
	&= \inf_{\sigma\in \ca{D}(\ca{H}_E)} \frac{1}{t}\log \sum_{x^n\in \ca{X}^n}Q_{X^n}(x^n) \prod_{i=1}^n \Xi_{1+t}(W(x_i)\|\sigma) \Label{eq:appnf3}\\
	& \le \inf_{\sigma\in \ca{D}(\ca{H}_E)} \sup_{x\in\ca{X}} \frac{1}{t}\log \left[ \Xi_{1+t}(W(x)\|\sigma) \right]^n \notag \\
	&= \inf_{\sigma\in \ca{D}(\ca{H}_E)} \sup_{x\in\ca{X}} n \tilde{D}_{1+t}(W(x)\|\sigma) \Label{eq:appnf5}\\
	&= n \sup_{Q_X} \tilde{I}_{1+t}^{\downarrow}(X;E|W\times Q_X) \Label{eq:appnf6},
\end{align}

where \eqref{eq:appnf1} and \eqref{eq:appnf5} follow from the definitions, \eqref{eq:appnf3} follows from the multiplicativity of $\Xi_{1+t}$ and \eqref{eq:appnf6} follows from Proposition \ref{lemma:P42}.
\end{proofof}

By using Lemma \ref{lemma:nf}, we have
\begin{align}
	\max_{Q_{X^n}}I_{1-t}(X^n;B^n|W_B^n\times Q_{X^n}) 
	&\ge  \max_{Q_{X^n}} I_{1-t}^{\downarrow}(X^n;B^n|W_B^n\times Q_{X^n}) \Label{eq:nfperror1}\\
	&= n \max_{Q_X} I_{1-t}^{\downarrow}(X;B|W_B \times Q_X), \Label{eq:nfperror2}
\end{align}
where \eqref{eq:nfperror1} follows from definition. \eqref{eq:nferror} is obtained by substituting above inequality into \eqref{eq:error}.
Hence, we obtain Corollary \ref{coro1}.

Combining \eqref{eq:DefIG} and \eqref{eq:ProIG}, we obtain the following bound
\begin{equation}
	\begin{aligned} \Label{eq:StrConvE}
		\tilde{I}_{1+t}^{\downarrow}(X;E|W_E\times P_{\ca{L}}) \le n \max_{Q_X} \tilde{I}_{1+t}^{\downarrow}(X;E|W_E \times Q_X)
	\end{aligned}
\end{equation}
where $P_{\ca{L}}$ is a distribution on $\ca{X}^n$ such that for $x^n\in \ca{L}$, $P_{\ca{L}}(x^n)=\frac{1}{\card{L}}$. Substituting \eqref{eq:StrConvE} into \eqref{eq:Trd} yields \eqref{eq:nfTD}. 
Hence, we obtain Corollary \ref{coro1-2}.

\subsection{Proof of Theorem \ref{thm:add}}\Label{A-thm:add}
For a given distribution $Q_X$ on $\ca{X}$, we have
\begin{align}
2^{t \tilde{I}_{1+t}^{\downarrow}(X;E|W_E\times Q_X)} = \inf_{\sigma \in \ca{D}(\ca{H}_E)}\sum_{x\in \ca{X}} Q_X(x) \Xi_{1+t}(W_{E}(x)\|\sigma) .
\end{align}
Given two distributions $Q_X$ and $\bar{Q}_X$ and $0<\lambda<1$, we have
\begin{align}
& 2^{t\tilde{I}_{1+t}^{\downarrow}(X;E|W_E\times (\lambda Q_X+(1-\lambda)\bar{Q}_X )} 
= \min_{\sigma \in \ca{D}(\ca{H}_E)}
\Big(\sum_{x\in \ca{X}} (\lambda Q_X(x)+(1-\lambda)\bar{Q}_X(x))  
\Xi_{1+t}(W_{E}(x)\|\sigma) \Big)\nonumber \\
\ge &
\lambda \min_{\sigma \in \ca{D}(\ca{H}_E)}\Big(\sum_{x\in \ca{X}} Q_X(x)  
\Xi_{1+t}(W_{E}(x)\|\sigma) \Big)+
(1-\lambda)\min_{\sigma \in \ca{D}(\ca{H}_E)}
\Big(\sum_{x\in \ca{X}} \bar{Q}_X(x)  \Xi_{1+t}(W_{E}(x)\|\sigma) \Big)\nonumber\\
=& \lambda 2^{t\tilde{I}_{1+t}^{\downarrow}(X;E|W_E\times  Q_X})
+ (1-\lambda)2^{t\tilde{I}_{1+t}^{\downarrow}(X;E|W_E\times \bar{Q}_X )} ,\Label{eq:addp3}
\end{align}
which implies that the map $Q_X\mapsto 
2^{t\tilde{I}_{1+t}^{\downarrow}(X;E|W_E\times Q_X)} $ is concave.

Given an element $x_0\in \ca{X}=G$, 
we define the distribution $Q_{X,x_0}$ and the cq channels
 $W_{E,x_0}$ and $ U_{x_0} (W_{E})$
as
$Q_{X,x_0}(x):=Q_X(x_0x)$, $W_{E,x_0}(x):=W_{E,x}(x_0 x)$, 
$ U_{x_0} \circ W_{E})(x):= U_{x_0} W_{E}(x) U_{x_0}^\dagger $,
respectively.
Then, we have
\begin{align}
&\tilde{I}_{1+t}^{\downarrow}(X;E|W_E \times Q_{X}) 
= \tilde{I}_{1+t}^{\downarrow}(X;E|W_{E,x_0} \times Q_{X,x_0}) \nonumber\\
=& \tilde{I}_{1+t}^{\downarrow}(X;E| U_{x_0} \circ W_{E} \times Q_{X,x_0})
= \tilde{I}_{1+t}^{\downarrow}(X;E|W_{E} \times Q_{X,x_0}), \Label{eq:addp0}
\end{align}
 where the second equality follows from the unitary invariance of quasi-entropy 
and the last inequality follows from the symmetric condition $W_E(x) = U_x \rho_0 U_x^\dagger$. 
Due to \eqref{eq:addp0} and \eqref{eq:addp3}, the uniform distribution
$P_{\ca{X}}$ on $\ca{X}$ satisfies 
\begin{align}
2^{t\tilde{I}_{1+t}^{\downarrow}(X;E|W_E \times Q_X )} 
&= \frac{1}{\card{X}} \sum_{x_0\in \ca{X}} 
2^{t\tilde{I}_{1+t}^{\downarrow}(X;E|W_E \times Q_{X,x_0})}\nonumber\\
&\le 2^{t\tilde{I}_{1+t}^{\downarrow}(X;E|W_E \times P_\ca{X})}. \Label{eq:addp4}
\end{align}
Thus, we have
\begin{equation}
\max_{P_X} \tilde{I}_{1+t}^{\downarrow}(X;E|W_E \times P_X)
=\tilde{I}_{1+t}^{\downarrow}(X;E|W_E \times P_\ca{X}).\Label{NLG}
\end{equation}
The combination of \eqref{eq:Trd} and \eqref{NLG} implies \eqref{eq:Trdadd}.

\subsection{Proof of Lemma \ref{A7}}\Label{A-A7}
It is known that the capacity of the degraded wire-tap channel 
is $\sup_{Q_X} I(X;{\hat{B}})_{Q_X}-I(X;{\hat{E}})_{Q_X} $ \cite{Devetak-Shor}, 
\cite[(9.75)]{hayashi2016quantum}.
Also, in this case,
For $x_0 \in \ca{X}$, we define the distribution $Q_{X,x_0}$ as
$Q_{X,x_0}(x):=Q_{X}(x_0 x)$.
Due to the symmetric condition, we find that
\begin{align}
I(X;{\hat{B}})_{Q_X}-I(X;{\hat{E}})_{Q_X} 
=I(X;{\hat{B}})_{Q_{X,x_0}}-I(X;{\hat{E}})_{Q_{X,x_0}}. 
\end{align}
Since $\sum_{x_0\in \ca{X}} \frac{1}{|\ca{X}|}Q_{X,x_0}= P_{\ca{X}}$
and the map $Q_X \mapsto I(X;{\hat{B}})_{Q_X}-I(X;{\hat{E}})_{Q_X} $
is known to be concave for degraded channels \cite[(9.76)]{hayashi2016quantum},
we have
\begin{align*}
I(X;{\hat{B}})_{P_{\ca{X}}}-I(X;{\hat{E}})_{P_{\ca{X}}} 
\ge & \sum_{x_0\in \ca{X}} \frac{1}{|\ca{X}|}\Big(
I(X;{\hat{B}})_{Q_{X,x_0}}-I(X;{\hat{E}})_{Q_{X,x_0}}\Big) \\
= &
I(X;{\hat{B}})_{Q_X}-I(X;{\hat{E}})_{Q_X} .
\end{align*}
Hence, 
$\sup_{Q_X} I(X;{\hat{B}})_{Q_X}-I(X;{\hat{E}})_{Q_X} =
I(X;{\hat{B}})_{P_{\ca{X}}}-I(X;{\hat{E}})_{P_{\ca{X}}} $.
We obtain Lemma \ref{A7}.

\section{Proof of Lemma \ref{LNM}}\Label{ANL}
 Recall the definition of $C$,
\begin{align}
	c = g_{s'}(m,y) = y + T(s')m.
\end{align}
For fixed $s',m$, the map between $y$ and $c$ is bijective so we have the function $y=y(m,s',c)$. 
Then, we evaluate $d(M; E' S' C )$ as follows.
\begin{align*}
&d(M; EA S' C )
=\min_{\sigma_{EA S' C}}
\| \tau_{M EA S' C}- P_{\ca{M}} \otimes \sigma_{EA S' C}\|_1 \\
\le &
\min_{ \{\sigma_{EA,Y=y,M=m}\}_{m,y}}
\Bigg\|
\sum_{s' ,m,y}
P_{\ca{M}}(m)P_{\ca{S}'}(s') P_{\ca{Y}}(y)
|m\rangle \langle m| 
\\
&\quad 
\otimes \tau_{EA,Y=y,M=m} 
\otimes |s'\rangle \langle s'| 
\otimes | g_{s'}(m,y )\rangle \langle g_{s'}(m,y )| \\
&-
\Big(\sum_{m} P_{\ca{M}}(m) 
|m\rangle \langle m| \Big)
\\
&\quad 
\otimes 
\Big(\sum_{s' ,m,y}P_{\ca{M}}(m) 
P_{\ca{S}'}(s') P_{\ca{Y}}(y)
\sigma_{EA,Y=y,M=m}
\otimes |s'\rangle \langle s'| 
\otimes | g_{s'}(m,y )\rangle \langle g_{s'}(m,y )| \Big)
\Bigg\|_1 \\
=&
\min_{ \{\sigma_{EA,Y=y,M=m}\}_{m,y}}
\Bigg\|
\sum_{s' ,m,c}
P_{\ca{M}}(m) P_{\ca{S}'}(s') P_{\ca{Y}}(c)
|m\rangle \langle m| \otimes \tau_{EA,Y=y(m,s',c),M=m}
\otimes |s'\rangle \langle s'| 
\otimes | c \rangle \langle c| \\
&-
\Big(\sum_{m} P_{\ca{M}}(m) 
|m\rangle \langle m| \Big)\otimes 
\Big(\sum_{s' ,m,c}P_{\ca{M}}(m) 
P_{\ca{S}'}(s')
P_{\ca{Y}}(c)
\sigma_{EA,Y=y(m,s',c),M=m}
\otimes |s'\rangle \langle s'| 
\otimes | c\rangle \langle c| \Big)
\Bigg\|_1 \\
=&
\min_{ \{\sigma_{EA,Y=y,M=m}\}_{m,y}}
\sum_{s' ,c}
P_{\ca{S}'}(s')
P_{\ca{Y}}(c)
\Bigg\|
\sum_{m}
P_{\ca{M}}(m) 
|m\rangle \langle m| \otimes \tau_{EA,Y=y(m,s',c),M=m}
\\
&-
\Big(\sum_{m} P_{\ca{M}}(m) 
|m\rangle \langle m| \Big)\otimes 
\Big(\sum_{m}P_{\ca{M}}(m) 
\sigma_{EA,Y=y(m,s',c),M=m}
\Big)
\Bigg\|_1 \\
=&
\min_{ \{\sigma_{EA,Y=y,M=m}\}_{m,y}}
\sum_{s' ,y}
P_{\ca{S}'}(s')
P_{\ca{Y}}(y)
\Bigg\|
\sum_{m}
P_{\ca{M}}(m) 
|m\rangle \langle m| \otimes \tau_{EA,Y=y,M=m}
\\
&-
\Big(\sum_{m} P_{\ca{M}}(m) 
|m\rangle \langle m| \Big)\otimes 
\Big(\sum_{m}P_{\ca{M}}(m) 
\sigma_{EA,Y=y,M=m}
\Big)
\Bigg\|_1 \\
=&
\min_{ \{\sigma_{EA,Y=y,M=m}\}_{m,y}}
\sum_{s'}
P_{\ca{S}'}(s')
\Bigg\|
\sum_{m,y}
P_{\ca{Y}}(y)
P_{\ca{M}}(m) 
|m,y\rangle \langle m,y| \otimes \tau_{EA,Y=y,M=m}
\\
&-
\sum_{y} P_{\ca{Y}}(y) 
\Big(\sum_{m} P_{\ca{M}}(m) 
|m,y\rangle \langle m,y| \Big)\otimes 
\Big(\sum_{m}P_{\ca{M}}(m) 
\sigma_{EA,Y=y,M=m}
\Big)
\Bigg\|_1 \\
\end{align*}
\begin{align*}
\le & \min_{ \{\sigma_{EA,M=m}\}_m}
\sum_{s'}
P_{\ca{S}'}(s')
\Bigg\|
\sum_{m,y}
P_{\ca{Y}}(y)
P_{\ca{M}}(m) 
|m,y\rangle \langle m,y| \otimes \tau_{EA,Y=y,M=m}
\\
&-
\sum_{y} P_{\ca{Y}}(y) 
\Big(\sum_{m} P_{\ca{M}}(m) 
|m,y\rangle \langle m,y| \Big)\otimes 
\Big(\sum_{m}P_{\ca{M}}(m) 
\sigma_{EA,M=m}
\Big)
\Bigg\|_1 \\
= & \min_{ \{\sigma_{EA,M=m}\}_{m}}
\sum_{s'}
P_{\ca{S}'}(s')
\Bigg\|
\sum_{m,y}
P_{\ca{Y}}(y)
P_{\ca{M}}(m) 
|m,y\rangle \langle m,y| \otimes \tau_{EA,Y=y,M=m}
\\
&-
\Big(\sum_{m,y} 
P_{\ca{M}}(m) P_{\ca{Y}}(y) 
|m,y\rangle \langle m,y| \Big)\otimes 
\Big(\sum_{m}P_{\ca{M}}(m) 
\sigma_{EA,M=m}
\Big)
\Bigg\|_1 \\
= & \min_{ \sigma_{EA}}
\sum_{s'}
P_{\ca{S}'}(s')
\Bigg\|
\sum_{m,y}
P_{\ca{Y}}(y)
P_{\ca{M}}(m) 
|m,y\rangle \langle m,y| \otimes \tau_{EA,Y=y,M=m}
\\
&-
\Big(\sum_{m,y} 
P_{\ca{M}}(m) P_{\ca{Y}}(y) 
|m,y\rangle \langle m,y| \Big)\otimes 
\sigma_{EA}
\Bigg\|_1 \\
=&d(M'; EA ).
\end{align*}
Then, we obtain \eqref{BMA}.

\section{Proof of Lemma \ref{lemma:AP}}\Label{A-lemma:AP}
Denote 
\begin{align}
	\omega_{XAE} &= W_E \times P_\ca{X} = \sum_{x\in \ca{X}} \frac{1}{\card{X}} \ket{x}\bra{x} \otimes U_x \tau_{AE} U_x^\dagger.
\end{align}
By definition,
\begin{align}
	\tilde{I}_{1+t}^{\downarrow} (X;AE|\omega_{XAE}) &= \min_{\sigma_{AE}} \frac{1}{t} \log \Xi_{1+t} (\omega_{XAE}\|\omega_X\otimes \sigma_{AE}). 
\end{align}
Notice that $\omega_{XAE}$ is invariant under group operation $\{U_x\}_{x\in\ca{X}}$ and quasi-entropy $\Xi_{1+t}(\cdot\|\cdot)$ (see \eqref{eq:DefQuasi} for definition) is invariant under unitary operation. Thus the minimizer $\sigma_{AE}$ is also invariant under group operation $\{U_x\}_{x\in\ca{X}}$, which means it takes the form $\sigma_{AE} = \frac{I_A}{d_A}$. Then we have
\begin{align}
	&\tilde{I}_{1+t}^{\downarrow} (X;AE|\omega_{XAE}) \nonumber\\
	&= \min_{\sigma_{E}} \frac{1}{t} \log \Xi_{1+t} (\omega_{XAE}\|\omega_X\otimes \frac{I_A}{d_A}\otimes \sigma_E) \nonumber\\ 
	&= \min_{\sigma_{E}} \frac{1}{t} \log \frac{d_A^s}{\card{X}} \sum_x \Xi_{1+t}(U_x \tau_{AE}U_x^\dagger\|\sigma_E) \nonumber\\
	&= \log d_A - \tilde{H}_{1+t}^\uparrow(A|E|\tau_{AE}).
\end{align}

\section{Proof of Theorem \ref{lemma:APB}}\Label{MMLA}
 Recall that an achievable rate for wire-tap channel requires 
$\epsilon(W_B) = \Pr[M\neq \hat{M}]$ and $\bar{d}(M;E)$ go to zero asymptotically. 
An achievable rate for PDC protocol requires 
$\epsilon_C, \epsilon_E$ and $\epsilon_B$ go to zero asymptotically. 
We will prove Theorem \ref{lemma:APB} by showing the conversion between these conditions.

Protocol 4 gives a conversion from a specific wire-tap code to 
a PDC protocol.
However, this type of conversion can be made for any 
wire-tap code.
Since the consumed length for covering variable $Y$
is negligible in comparison with $n$, we have the relation $\le$ in \eqref{XL1}.

Conversely, the asymptotic setting of the PDC model requires the following;
Alice transmits her message with asymptotically zero error
under $n$ times use of the channel $W_B$.
Also, when Eve receives her information via 
$n$ times use of the channel $W_E$, Eve 
obtains no information for Alice's message. 

Now, we consider a PDC protocol $P_n$
of $\epsilon_C-$ complete, $\epsilon_E-$ secure, and $\epsilon_B-$ reliable
with a message $M\in \ca{M}$ and a random variable 
$S\in \ca{S}$ to be sent via public channel.
We have a conditional distribution $P_{G^n|M,S}$ such that
Alice chooses an element $(g_1, \ldots, g_n)\in G^n$ subject to $P_{G^n|M,S}$
and applies $U_{g_1}\otimes\cdots \otimes U_{g_n}$
dependently of $S$ and $M$.
Also, we denote the decoder $\Pi^S$ dependently of $S$.

In the following, we consider the case when 
the channel from Alice to Bob is $\Lambda_A^{\otimes n}$.
We denote the recovered message by Bob by $\hat{M}$.
Then, we have
\begin{align}
\Pr[M\neq \hat{M}] &= \Pr[M \neq \hat{M}, \Ab^c] + \Pr[M\neq \hat{M}, \Ab]\\
&\le \Pr[\Ab^c | M\neq \hat{M}] + \Pr[\Ab] \\
&\le \epsilon_B+\epsilon_C,
\end{align}
Also, there exists a distribution $Q_S$ of $S$ such that
\begin{align}
\| P_{S,M}-P_M\times Q_S \|_1\le \epsilon_E,
\end{align}
which implies
\begin{align}
\| P_{S}-Q_S \|_1\le \epsilon_E.
\end{align}
Thus, we have
\begin{align}
\| P_{S,M}-P_M\times P_S \|_1\le  2 \epsilon_E.\Label{NLA}
\end{align}
Hence, we have
\begin{align}
\sum_{s \in \ca{S}}P_S(s) \Big(
P(M\neq \hat{M}|S=s)
+\| P_{M|S=s}-P_M \|_1
\Big)
\le 2 \epsilon_E+ \epsilon_B+\epsilon_C.
\end{align}
When $S=s$ and the distribution $P_{M|S=s}$ is replaced by $P_M$
we denote the decoding error probability
by $\Pr[M\neq \hat{M}|S=s][P_M] $.
Then, we have
\begin{align}
\Pr[M\neq \hat{M}|S=s][P_M]
\le &
\Pr[M\neq \hat{M}|S=s][P_{M|S=s}]
+\| P_{M|S=s}-P_M \|_1 \nonumber\\
=&
\Pr[M\neq \hat{M}|S=s]
+\| P_{M|S=s}-P_M \|_1.
\end{align}
That is, we have
\begin{align}
\sum_{s \in \ca{S}}P_S(s) 
\Pr[M\neq \hat{M}|S=s][P_M]
\le 2 \epsilon_E+ \epsilon_B+\epsilon_C.\Label{CL1}
\end{align}

In the following, we consider the case when 
Eve intercepts the transmitted state.
There exists a state $\sigma_{SE}$
\begin{align}
\| \rho_{MSE}- P_M \otimes \sigma_{SE}\|\le \epsilon_E.
\end{align}
Due to the same reason as \eqref{NLA}, we have
\begin{align}
\| \rho_{MSE}- P_M \otimes \rho_{SE}\|\le 2 \epsilon_E.
\end{align}
We define the joint state $\bar{\rho}_{MSE}$ as
\begin{align}
\bar{\rho}_{MSE}:=
\sum_{s \in \ca{S},m\in \ca{M}}P_M(m)P_S(s)
|m,s\rangle \langle m,s| \otimes \rho_{E| M=m,S=s}.
\end{align}
Then, we have
\begin{align}
\| \bar{\rho}_{MSE} - P_M \otimes \rho_{SE} \|_1
\le &
\| \bar{\rho}_{MSE} -{\rho}_{MSE} \|_1
+\| {\rho}_{MSE} - P_M \otimes \rho_{SE} \|_1 \nonumber\\
\le &
\| P_{S,M}-P_M\times P_S \|_1
+\| {\rho}_{MSE} - P_M \otimes \rho_{SE} \|_1 
\le 4 \epsilon_E.
\end{align}
That is,
\begin{align}
\sum_{s \in \ca{S}}P_S(s) 
\| \bar{\rho}_{ME|S=s} - P_M \otimes \rho_{E|S=s} \|_1
\le 4 \epsilon_E.\Label{CL2}
\end{align}
Due to \eqref{CL1}, \eqref{CL2}, and Markov inequality,
there exists $s_0 \in \ca{S}$ such that
\begin{align}
P(M\neq \hat{M}|S=s_0)[P_M]
&\le 3(2 \epsilon_E+ \epsilon_B+\epsilon_C) \Label{CL3}\\
\| \bar{\rho}_{ME|S=s_0} - P_M \otimes \rho_{E|S=s_0} \|_1
&\le 12 \epsilon_E.
\end{align}
In the same way as \eqref{NLA}, we have
\begin{align}
\| \bar{\rho}_{ME|S=s_0} - P_M \otimes \bar{\rho}_{E|S=s_0} \|_1
&\le 24 \epsilon_E.\Label{CL4}
\end{align}

Now, as our wire-tap encoder,
we choose $P_{G^n|M=m,S=s_0}$ for each message $m \in \ca{M}$.
We choose $\Pi^{s_0}$ as our wire-tap decoder.
The decoding error probability of this wire-tap code is evaluated by \eqref{CL3},
and 
the information leakage of this wire-tap code is evaluated by \eqref{CL4}.
Hence, we have the relation $\ge$ in \eqref{XL1}.

\bibliographystyle{unsrt}
\bibliography{References.bib}

\end{document}